\documentclass[aps,prd,preprint,groupedaddress,showkeys,showpacs,notitlepage]{revtex4-2}
\usepackage{graphicx,epsfig,dcolumn,bm,epic,eepic,float}
\usepackage{amsmath}
\usepackage{hyperref}
\usepackage{latexsym}
\usepackage{float}
\usepackage{graphicx, caption}
\usepackage{listings}
\usepackage[dvipsnames]{xcolor}
\usepackage{booktabs}
\usepackage{hyperref}
\usepackage[noabbrev]{cleveref}
\usepackage{makeidx,shortvrb,latexsym}  
\usepackage{hyperref}

\newcommand\YAMLcolonstyle{\color{red}\mdseries}
\newcommand\YAMLkeystyle{\color{black}\bfseries}
\newcommand\YAMLvaluestyle{\color{blue}\mdseries}

\makeatletter

\newcommand\language@yaml{yaml}

\expandafter\expandafter\expandafter\lstdefinelanguage
\expandafter{\language@yaml}
{
  keywords={true,false,null,y,n},
  keywordstyle=\color{darkgray}\bfseries,
  basicstyle=\YAMLkeystyle,                                 
  sensitive=false,
  comment=[l]{\#},
  morecomment=[s]{/*}{*/},
  commentstyle=\color{purple}\ttfamily,
  stringstyle=\YAMLvaluestyle\ttfamily,
  moredelim=[l][\color{orange}]{\&},
  moredelim=[l][\color{magenta}]{*},
  moredelim=**[il][\YAMLcolonstyle{:}\YAMLvaluestyle]{:},   
  morestring=[b]',
  morestring=[b]",
  literate =    {---}{{\ProcessThreeDashes}}3
                {>}{{\textcolor{red}\textgreater}}1     
                {|}{{\textcolor{red}\textbar}}1 
                {\ -\ }{{\mdseries\ -\ }}3,
}

\lst@AddToHook{EveryLine}{\ifx\lst@language\language@yaml\YAMLkeystyle\fi}
\makeatother

\newcommand\ProcessThreeDashes{\llap{\color{cyan}\mdseries-{-}-}}

\lstdefinestyle{cpp}{
	language=C++,
    basicstyle=\ttfamily\scriptsize,
	keywordstyle=\color{blue},
	commentstyle=\color{green},
	stringstyle=\color{red},
	showstringspaces=false,
	numbers=none,
	tabsize=2,	
}

\lstdefinestyle{bash}{
    language=bash,
    basicstyle=\ttfamily\scriptsize,
    keywordstyle=\color{blue},
    commentstyle=\color{green},
    stringstyle=\color{red},
    showstringspaces=false,
    numbers=none,
    tabsize=2,
}

\lstdefinestyle{CMake}{
    language=bash,
    basicstyle=\ttfamily\scriptsize,
    keywordstyle=\color{blue},
    commentstyle=\color{green},
    stringstyle=\color{red},
    showstringspaces=false,
    numbers=none,
    tabsize=2,
}

\begin{document}
\unitlength 1 cm

\title{PDFxTMDLib: A High-Performance C++ Library for Collinear and Transverse Momentum Dependent Parton Distribution Functions }

\author{R. Kord Valeshabadi}
\altaffiliation {Corresponding author, Email:
ramin.kord@ipm.ir}
\author{S.~Rezaie}
\affiliation{School of Particles and Accelerators, Institute for Research in Fundamental Sciences (IPM), P. O. Box      19395-5531, Tehran, Iran\\}

\date{\today}
\begin{abstract}
Collinear parton distribution functions (cPDFs) and transverse momentum dependent distributions (TMDs) are essential for calculating cross sections in high-energy physics, particularly within collinear and $k_t$-factorization frameworks. Currently, there exists two libraries, such as \texttt{LHAPDF} and \texttt{TMDLib}, to obtain these physical objects. However, there are limitations in both libraries, especially for TMDs, such as restricted customization and extensibility. Users are limited to the implementations provided by these libraries and cannot easily support unconventional PDFs. Moreover, no standard TMD library currently provides a consistent framework for QCD coupling evaluation or for studying uncertainties at the distribution level—features that are important for diagnostic and comparative analyses in phenomenological research.

To address these shortcomings, we introduce \texttt{PDFxTMDLib}, a modern C++ library designed to offer a robust and flexible solution. This library supports both collinear PDFs and TMDs while allowing greater customization. It also opens the way to support higher-order distributions. In this article, we describe the structure of \texttt{PDFxTMDLib}. We also demonstrate its validity and performance by integrating it into the PYTHIA Monte Carlo event generator to compute Drell-Yan cross sections. Additionally, comparisons of PDFs obtained from \texttt{PDFxTMDLib} with those from \texttt{LHAPDF} and \texttt{TMDLib} confirm the reliability of \texttt{PDFxTMDLib}'s results.
	\end{abstract}
\pacs{12.38.Bx, 13.85.Qk, 13.60.-r
	\\ \textbf{Keywords:}  PDFxTMDLib, Collinear Parton Distribution Functions, Transverse Momentum Parton Distribution Functions, PDFs, TMDs, UPDFs, Collinear factorization, $k_t$-factorization, Interpolation Library, LHAPDF, TMDLib} \maketitle
\newpage
\section{Introduction}
Parton distribution functions (PDFs) are fundamental quantities in high energy physics that characterize the momentum distribution of partons (quarks and gluons) within hadrons. These distributions play a crucial role in predicting cross sections for high energy collisions at facilities such as the Large Hadron Collider (LHC). Two main theoretical frameworks have been developed for calculating hadronic cross sections: collinear factorization and $k_t$-factorization. In collinear factorization, partons are assumed to move parallel to the hadron's momentum direction, neglecting transverse motion. This framework relies on collinear PDFs (cPDFs), which depend only on the longitudinal momentum fraction $x$ and factorization scale $\mu$. In contrast, $k_t$-factorization incorporates transverse momentum effects through TMDs or unintegrated parton distribution functions (UPDFs), which additionally depend on the transverse momentum $k_t$. This latter approach is particularly important for processes involving small-$x$ physics or high energy collisions.

In both frameworks, hadronic cross sections are expressed as convolutions of partonic cross sections and PDFs. For collinear factorization, the cross section takes the form:
\begin{equation}
\label{som}
\sigma  = \sum_{i,j \in {q,g}}\int \dfrac{dx_1}{x_1}\dfrac{dx_2}{x_2}f_i(x_1,\mu^2) f_j(x_2,\mu^2)\hat{\sigma}_{ij},
\end{equation}
where $f_{i(j)}$ are cPDFs depending on the longitudinal momentum fractions $x_{1,2}$ and the factorization scale $\mu$.

For $k_t$-factorization, the cross section has the more general form:
\begin{equation}
\label{som2}
\sigma  = \sum_{i,j \in {q,g}}\int \dfrac{dx_1}{x_1}\dfrac{dx_2}{x_2} \dfrac{dk_{1,t}^2}{k_{1,t}^2}\dfrac{dk_{2,t}^2}{k_{2,t}^2} f_i(x_1, k_{1,t}^2, \mu^2) f_j(x_2, k_{2,t}^2, \mu^2)\hat{\sigma}_{ij}^*,
\end{equation}
where $f_{i(j)}$ represents TMDs, which additionally depend on the transverse momenta $k_{1,t}$ and $k_{2,t}$ of the partons. The partonic cross section $\hat{\sigma}_{ij}^*$ is off-shell due to the transverse momenta of the incoming partons.

The evolution of cPDFs is governed by the Dokshitzer–Gribov–Lipatov–Altarelli–Parisi (DGLAP) evolution equations \cite{DGLAP1, DGLAP2, DGLAP3}, which describe the scale dependence of cPDFs within perturbative quantum chromodynamics (QCD). These equations account for logarithmic corrections from parton emissions, enabling the evolution of cPDFs from a specified initial scale to higher momentum transfers. The initial distributions are determined through global fits comparing theoretical predictions with experimental data from processes such as deep inelastic scattering (DIS), Drell-Yan production, and hadronic collisions \cite{Alekhin2017, NNPDF2015}.

In contrast, the $k_t$-factorization formalism employs evolution equations such as the Ciafaloni-Catani-Fiorani-Marchesini (CCFM)\cite{CCFM1, CCFM2, CCFM3} and Balitsky-Fadin-Kuraev-Lipatov (BFKL) \cite{BFKL1,BFKL2} equations, which govern the scale dependence of TMDs. However, these evolution equations are primarily limited to gluons, posing challenges in obtaining TMDs for all parton species. While CCFM-based TMDs have been extended to include valence quarks \cite{CCFMQ}, a complete set covering all quark flavors remains unavailable. Modern approaches such as parton branching (PB)\cite{PB1}, Kimber-Martin-Rysking (KMR) \cite{KMR}, and Martin-Ryskin-Watt (MRW)\cite{MRW} use DGLAP evolution equations to provide mechanisms for obtaining TMDs for both quarks and gluons, making cross section calculations in the $k_t$-factorization framework increasingly practical.

Generally, cPDFs and TMDs are provided in the form of grid files, where interpolation based libraries are used to calculate them. The \texttt{LHAPDF} library \cite{LHAPDF6} is widely used to access collinear Parton Distribution Functions (cPDFs) for calculating cross sections within the collinear factorization framework. Similarly, TMDLib \cite{TMDLib} is commonly employed for TMDs in cross section calculations within the $k_t$-factorization framework. These libraries primarily facilitate interpolation-based access, enabling efficient retrieval of cPDFs via two-dimensional interpolations and TMDs via three-dimensional interpolations.

However, the main limitations of these libraries are their lack of extensibility and portability. Their core design, based on fixed-dimensional interpolation, lacks the extensibility required for modern phenomenological studies involving higher-order distributions like Double Parton Distribution Functions (DPDFs), etc. Furthermore, users are confined to the libraries' built-in algorithms, with no straightforward way to implement custom interpolation or extrapolation methods.

For \texttt{TMDLib}, these issues manifest differently. Although common metadata
(info) files are available for all sets, there is no widely adopted standard for
TMD sets regarding the grid or file structure (and, in practice, the grid layout
conventions). Interpolation is performed according to the prescriptions defined
by the respective set authors, as described in the \texttt{TMDLib} manual. The
library is, in fact, a collection of independent components, each handling its
specific TMD set under a common Abstract Programming Interface (API). Due to
this lack of a standardized grid/file procedure, adding support for a new TMD set
remains a non-trivial task. Additionally, while uncertainty members (where
provided) are stored in a standardized way, uncertainty propagation cannot be
performed within the library itself at the level of the distributions and is only
feasible through \texttt{TMDplotter}~\cite{TMDLib}.  This tool enables
diagnostic analyses of PDF- or TMD-level uncertainties in a manner consistent with \texttt{LHAPDF}. In \texttt{PDFxTMDLib}, this functionality is extended by introducing an explicit \texttt{ErrorType} field in the
metadata, following the \texttt{LHAPDF} convention, to enable consistent handling of PDF- and TMD-level uncertainty information.

In this work, we introduce \texttt{PDFxTMDLib} (see \url{https://github.com/Raminkord92/PDFxTMD}), a comprehensive library designed to handle both cPDFs and TMDs within a unified framework. Beyond fully supporting \texttt{LHAPDF} features, it provides a standardized approach to TMD calculations, simplifying access to uncertainty quantification, correlation analysis, and QCD coupling determination. The library implements modern C++ design principles to achieve high performance while maintaining extensibility, resulting in a robust and efficient framework for PDF computations.

A key innovation of \texttt{PDFxTMDLib} is its architecture, which currently supports 2D and 3D interpolations but is explicitly designed to extend to higher-dimensional cases required for advanced distribution functions such as double parton distribution functions (DPDFs) and Double Transverse Momentum Dependent parton distribution functions (DTMDs). This forward-looking design enables researchers to conduct complex analyses as theoretical models evolve. Furthermore, the library's modular architecture allows users to implement custom components, like readers, interpolators, and extrapolators, suitable to their specific research requirements. This flexibility ensures that \texttt{PDFxTMDLib} can adapt to emerging theoretical developments and specialized applications in high energy physics. 

This article is organized as follows. Section \ref{architecture} presents the library architecture, the core interfaces for reading, interpolation, and extrapolation of PDF data. We describe different interfaces which allow users to calculate cPDFs, TMDs, QCD coupling, and uncertainty. Section \ref{validation} provides numerical validation and performance comparisons of \texttt{PDFxTMDLib} against \texttt{LHAPDF} and \texttt{TMDLib}. Section \ref{installationAndUsage} covers the installation process and provides detailed examples of how to use the library's API for both cPDFs and TMDs calculations. Finally, section \ref{conclusion} concludes with a discussion of the library's impact and potential future developments.
\section{Architecture of PDFxTMDLib}
\label{architecture}

This section presents a systematic analysis of the \texttt{PDFxTMDLib} architecture, focusing on its foundational design principles, key components, and interaction patterns. The architecture employs modern C++ idioms to achieve both computational efficiency and design flexibility. We utilize Unified Modeling Language (UML) diagrams to precisely illustrate component relationships and structural hierarchies within the system. It should be noted that in the UML diagrams representing classes with its members, we do not present all member variables and methods, and only show important ones. Users can check this documentation at \url{https://pdfxtmdlib.org/docs/} for further details.

\subsection{Architectural Overview and Core Design Principles}

\texttt{PDFxTMDLib} is architected around three fundamental design principles: (1) separation of concerns through clearly defined interfaces, (2) compile-time polymorphism for performance optimization, and (3) type erasure for interface flexibility. These principles allow the library to achieve both high performance and extensibility, characteristics that are often in tension in scientific computing libraries.

\begin{figure}[htbp]
\centering
\includegraphics[width=0.9\textwidth]{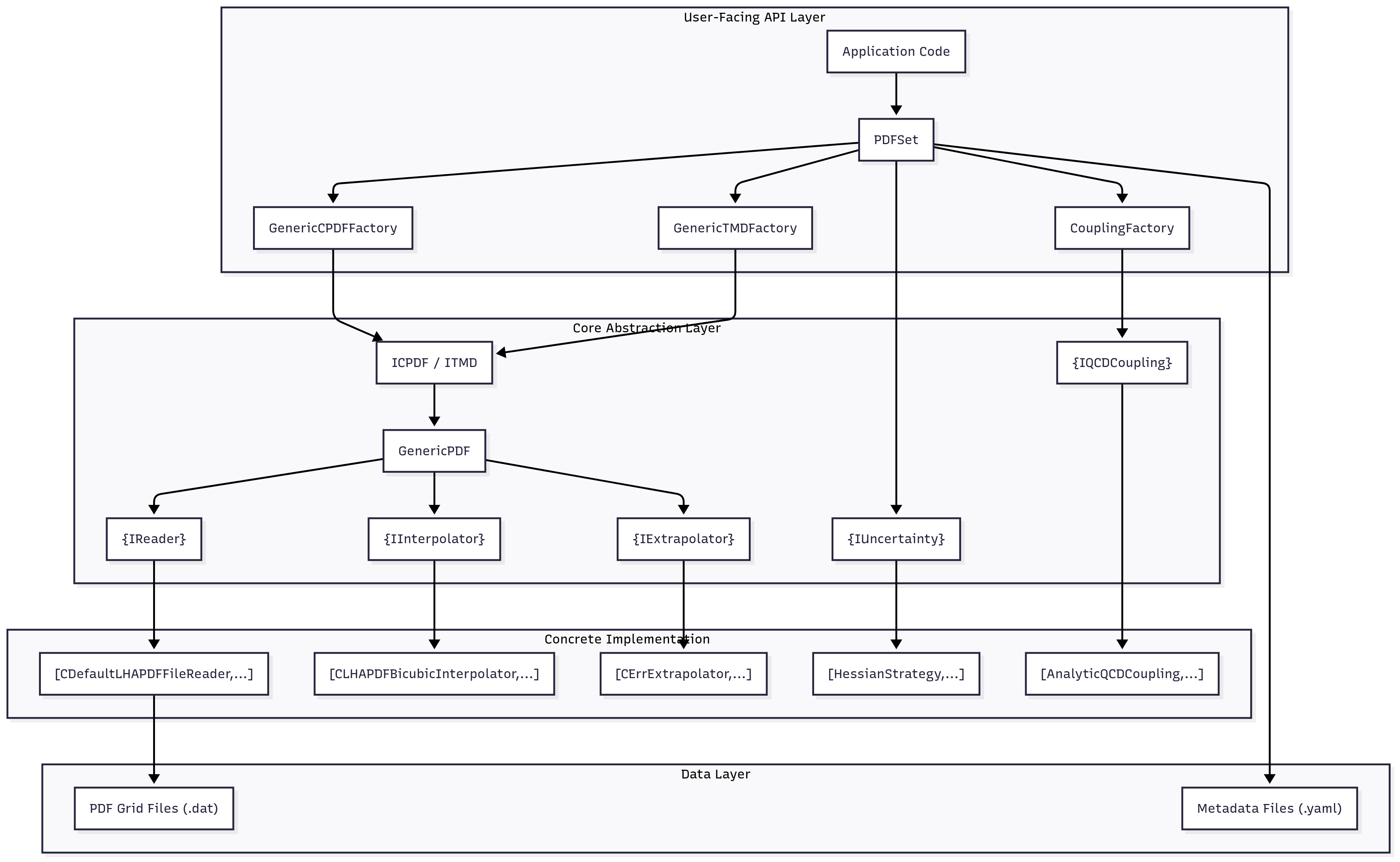}
\caption{UML diagram providing a general view of the \texttt{PDFSet} class and its interactions with other components in the \texttt{PDFxTMDLib} architecture.}
\label{fig:birdEyeViewUML}
\end{figure}

At the top of \texttt{PDFxTMDLib}'s architecture is the \texttt{PDFSet} class, which serves as the primary user-facing component. This template class provides a unified interface for computing various quantities including cPDFs, TMDs, QCD coupling constants, and associated statistical measures such as uncertainties and correlations. Figure~\ref{fig:birdEyeViewUML} illustrates the hierarchical structure of the codebase through a UML class diagram, highlighting the relationships between primary components. Before explaining diagram, one should be careful that \texttt{[CDefaultLHAPDFFileReader, ...]} means concrete classes such as \texttt{CDefaultLHAPDFFileReader} that implements the interface \texttt{IReader}.

The library's input mechanism accepts either cPDF or TMD sets, each comprising grid data with predefined dimensionality and an associated metadata file in YAML format. The \texttt{PDFSet} class delegates instantiation responsibilities to specialized factory classes, i.e. \texttt{GenericCPDFFactory}, \texttt{GenericTMDFactory}, and \texttt{CouplingFactory}, which create concrete objects implementing the \texttt{ICPDF}, \texttt{ITMD}, and \texttt{IQCDCoupling} interfaces, respectively. These factory classes parse the metadata file to determine the appropriate implementation strategies for each computational component. Notably, the factory-produced objects maintain implementation independence, allowing advanced users to utilize them directly without the \texttt{PDFSet} abstraction layer.

\subsection{Component Interactions}

\begin{figure}[htbp]
\centering
\includegraphics[width=0.4\textwidth]{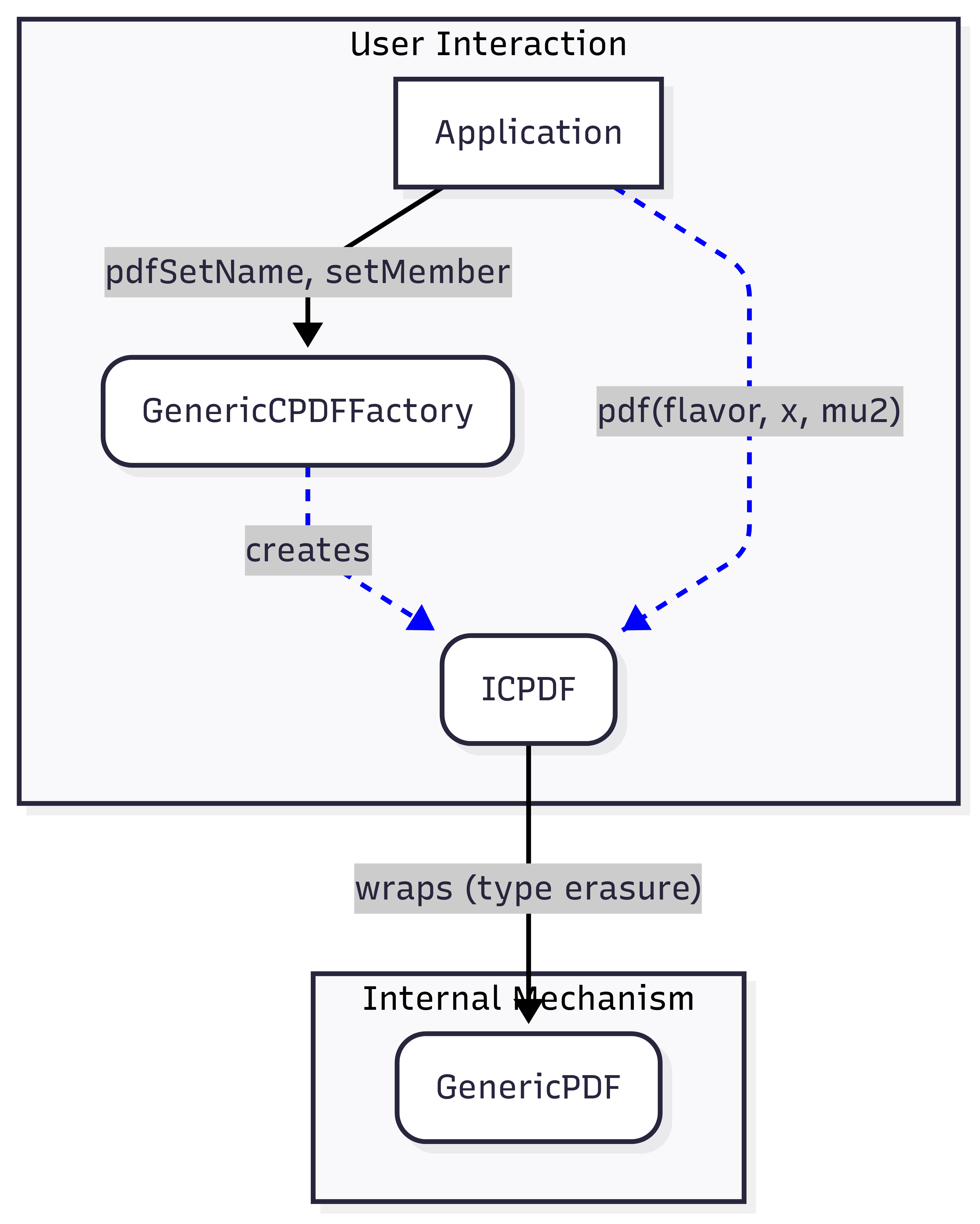}
\includegraphics[width=0.4\textwidth]{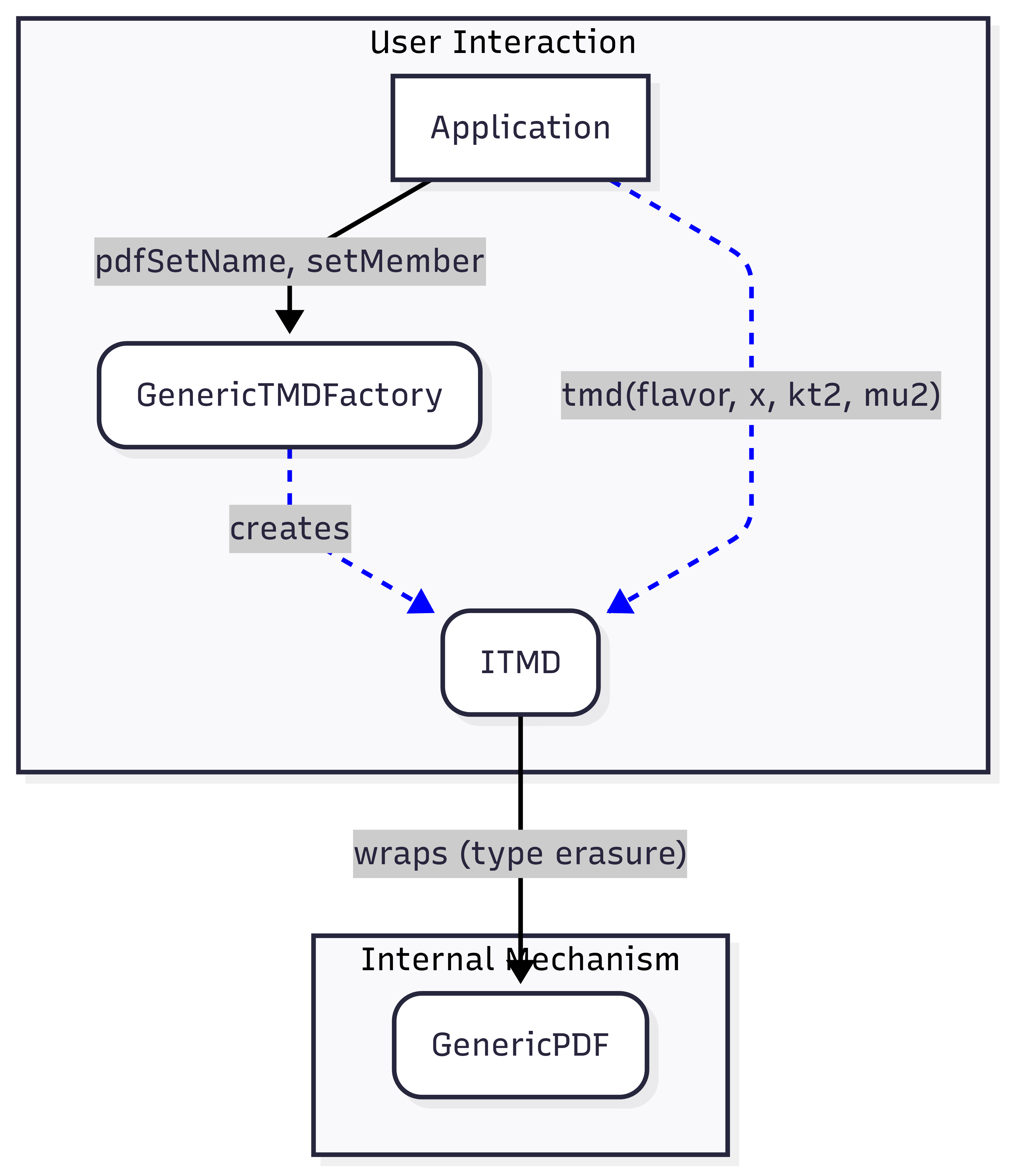}
\caption{UML diagrams illustrating the structures of the \texttt{GenericCPDFFactory} (left) and \texttt{GenericTMDFactory} (right) classes within the \texttt{PDFxTMDLib} architecture.}
\label{fig:GenericTMD-CPDF}
\end{figure}

Figures~\ref{fig:GenericTMD-CPDF} present UML diagrams depicting the structures of the \texttt{GenericCPDFFactory}\allowbreak and \texttt{GenericTMDFactory} classes, respectively. These factory classes are responsible for instantiating objects that implement the \texttt{ICPDF} and \texttt{ITMD} interfaces, essential for computing cPDFs and TMDs. The instantiation process is facilitated by the \texttt{GenericPDF} class, which encapsulates three critical components: the \texttt{IReader} interface for reading PDF set grid data, the \texttt{IInterpolator} interface for performing interpolation between grid points, and the \texttt{IExtrapolator} interface for extrapolation beyond grid boundaries. It should be noted that interfaces \texttt{IReader}, \texttt{IInterpolator} and \texttt{IExtrapolator} are implemented through CRTP design pattern to improve the performance, minimize performance overhead of using runtime polymorphism. 

For users requiring only cPDFs or TMDs, the \texttt{GenericCPDFFactory} and \texttt{GenericTMDFactory} classes offer direct access to these computations without necessitating the \texttt{PDFSet} class. Advanced users may further customize calculations by selecting specific implementations of the \texttt{IReader}, \texttt{IInterpolator}, and \texttt{IExtrapolator} interfaces. However, the \texttt{IUncertainty} interface, integral to uncertainty and correlation calculations, remains dependent on the \texttt{PDFSet} class, as it requires access to the PDF set members.

The subsequent subsections provide detailed insights into the \texttt{PDFSet}, factory, and \texttt{GenericPDF} classes, emphasizing their roles and interactions within the \texttt{PDFxTMDLib} framework.

\subsubsection{The \texttt{PDFSet} Class}
\label{sec:pdfset_class}

The \texttt{PDFSet} class serves as the primary interface for users interacting with a specific parton distribution function (PDF) set. It offers a high-level API for evaluating PDFs, computing uncertainties, determining correlations, and retrieving the QCD coupling constant (\(\alpha_s\)). While conceptually similar to the \texttt{PDFSet} class in the \texttt{LHAPDF} library, this implementation improves usability by unifying the interface for both cPDFs and TMDs, while simplifying uncertainty and correlation computations through dedicated methods.

\begin{figure}[htbp]
\centering
\includegraphics[width=\textwidth]{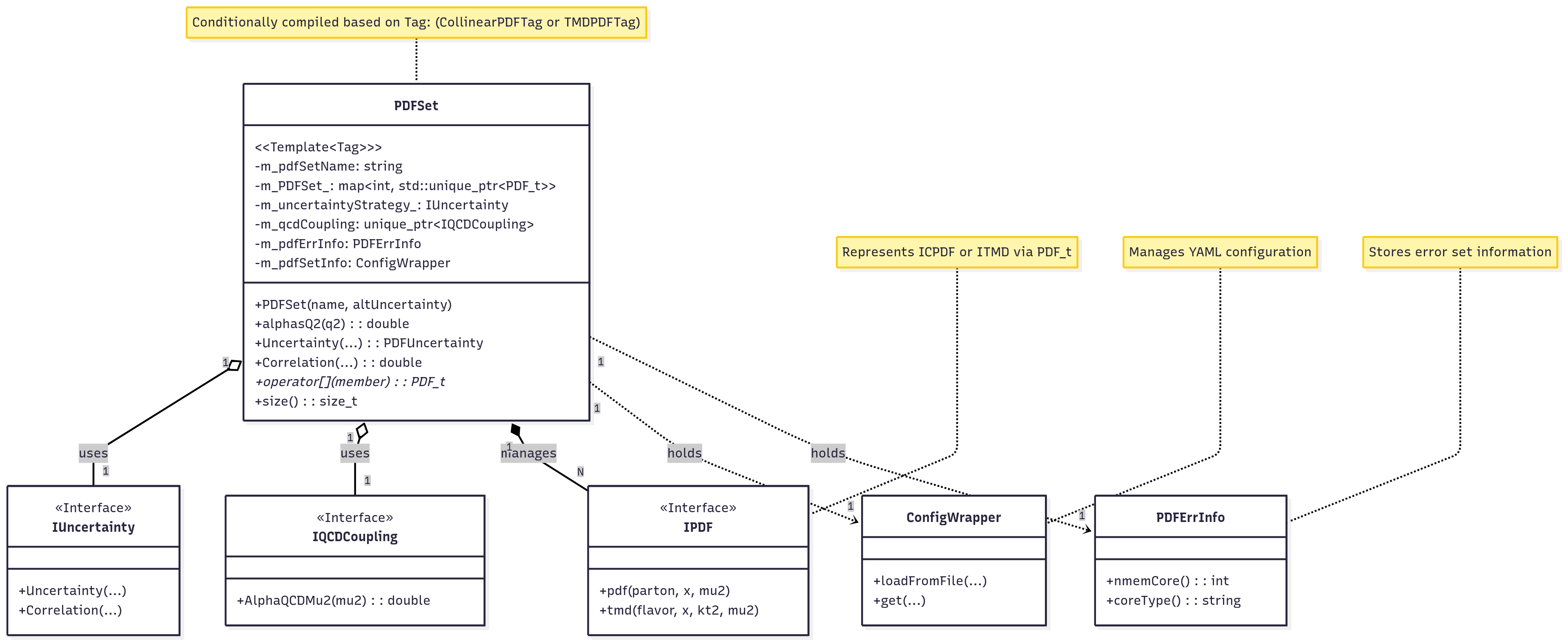}
\caption{UML diagram detailing the structure and interactions of the \texttt{PDFSet} class within the \texttt{PDFxTMDLib} architecture.}
\label{fig:PDFSet}
\end{figure}

Figure~\ref{fig:PDFSet} presents a UML diagram of the \texttt{PDFSet} class, illustrating its attributes, methods, and class dependencies. As a template class parameterized by \texttt{TAG} (either \texttt{CollinearPDFTag} or \texttt{TMDPDFTag}), it employs conditional compilation to adapt its behavior to either cPDFs or TMDs.

The key attributes of the class are:

\begin{itemize}
\item \texttt{m\_PDFSetName: std::string},  stores the name of the PDF set.
\item \texttt{m\_PDFSet: std::map<int, std::unique\_ptr<PDF\_t>>}, maintains a collection of PDF set members. Here, \texttt{PDF\_t} is an alias that resolves to either \texttt{ICPDF} or \texttt{ITMD}, depending on the \texttt{TAG}.
\item \texttt{m\_uncertaintyStrategy: IUncertainty}, provides functionality for uncertainty and correlation computations.
\item \texttt{m\_qcdCoupling: std::unique\_ptr<IQCDCoupling>}, used to evaluate the QCD running coupling \(\alpha_s\) at a given scale.
\item \texttt{m\_PDFErrInfo: PDFErrInfo}, stores error-related metadata, such as the index of the central member.
\item \texttt{m\_selfInfo: ConfigWrapper}, manages YAML-based metadata associated with the PDF set.
\end{itemize}

The class defines the following primary methods:

\begin{itemize}
\item A constructor for initializing the class with a specific PDF set.
\item \texttt{AlphaQCDMu2(mu2)}, Computes the strong coupling \(\alpha_s\) at a given scale \(\mu^2\).
\item \texttt{Uncertainty(...)} and \texttt{Correlation(...)}, Compute statistical uncertainties and correlations.
\item \texttt{operator[](member)}, Provides access to individual PDF set members via the specified index.
\item \texttt{size()}, Returns the number of available members in the PDF set.
\end{itemize}

The UML diagram in Figure~\ref{fig:PDFSet} highlights the class’s dependencies and relationships with key components: \texttt{IUncertainty} (for uncertainty calculations), \texttt{IQCDCoupling} (for computing \(\alpha_s\)), \texttt{ConfigWrapper} (for reading metadata), and \texttt{PDFErrInfo} (for managing error information).

This unified interface, enabled by the \texttt{TAG} parameter, supports flexible, physics-oriented computation suitable to both collinear and TMD use cases, which can also be extended to more tags regarding higher dimensional PDFs, which making \texttt{PDFxTMDLib} applicable to a broad range of high energy physics analyses.

\subsection{Factory Classes}

\begin{figure}[htbp]
\centering
\includegraphics[width=1\textwidth]{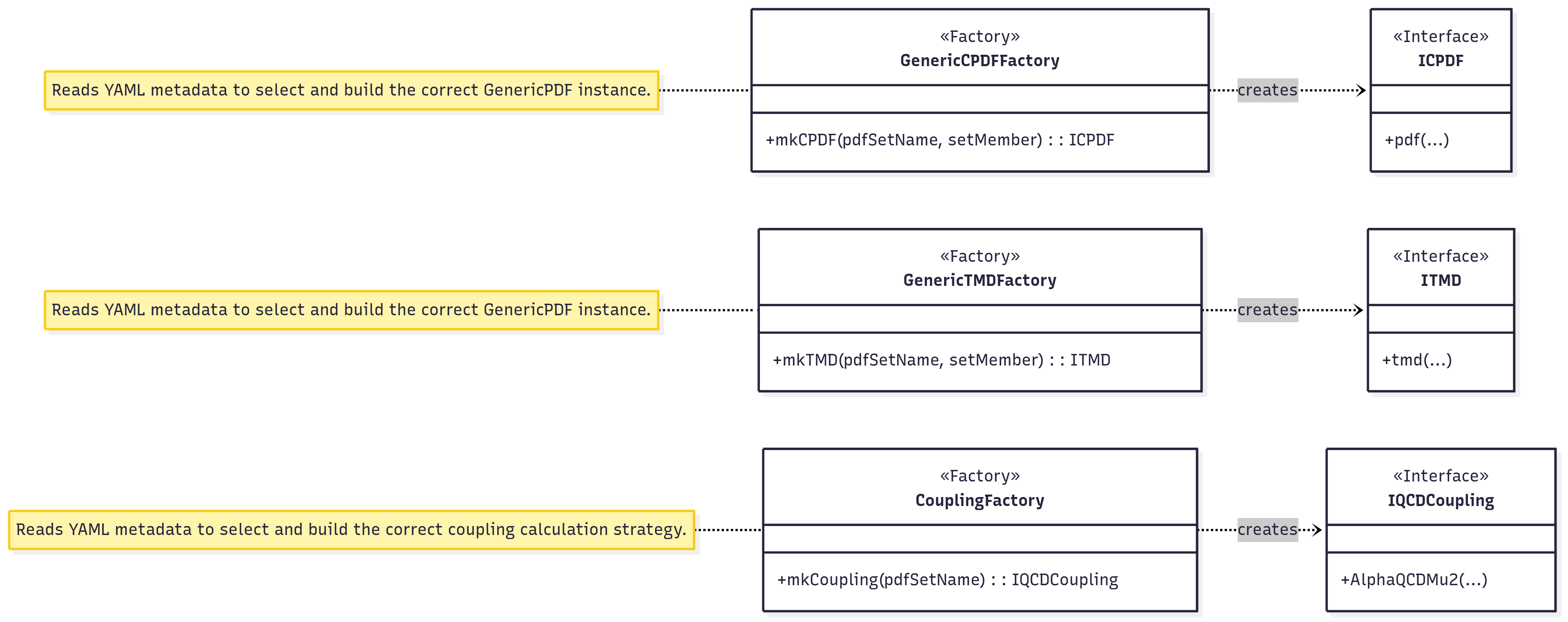}
\caption{UML diagram showcasing the interactions of factory classes (\texttt{GenericCPDFFactory}, \texttt{GenericTMDFactory}, \texttt{CouplingFactory}) within the \texttt{PDFxTMDLib} architecture.}
\label{fig:FactoryUML}
\end{figure}

The factory classes, i.e. \texttt{GenericCPDFFactory}\allowbreak, \texttt{GenericTMDFactory}\allowbreak, and \texttt{CouplingFactory}\allowbreak, facilitate the creation of \texttt{ICPDF}, \texttt{ITMD}, and \texttt{IQCDCoupling} objects, respectively. Figure~\ref{fig:FactoryUML}, provides a UML diagram showcasing their interactions. The \texttt{GenericCPDFFactory} ensures a method \texttt{mkCPDF(pdfSetName, setMember): ICPDF}, creating an \texttt{ICPDF} object for cPDF computation via \texttt{pdf(...)}, guided by YAML metadata to select the appropriate \texttt{GenericPDF} instance. Similarly, \texttt{GenericTMDFactory} offers \texttt{mkTMD(pdfSetName, setMember): ITMD} for TMDs via \texttt{tmd(...)}, using the same metadata-driven process. The \texttt{CouplingFactory} provides \texttt{mkCoupling(pdfSetName): IQCDCoupling} for $\alpha_s$ computation via \texttt{AlphaQCDMu2(...)}, selecting strategies from YAML metadata. These factories employ type erasure, which provides flexibility and type safety without traditional inheritance (see \cite{iglberger_cpp_design}). Using type erasure in fact highly improve performance, where in this work we use the technique of manual implementation of function dispatch, which essentially avoid runtime polymorphism.  This design allows advanced users to directly use factory-produced objects, bypassing PDFSet. The diagrams in Figures~\ref{fig:GenericTMD-CPDF} further illustrate these processes for \texttt{GenericCPDFFactory} and \texttt{GenericTMDFactory}, showing their creation of \texttt{ICPDF} and \texttt{ITMD} objects that wrap \texttt{GenericPDF} via type erasure, with users invoking computation methods directly.

\subsection{GenericPDF Class}

\begin{figure}[htbp]
\centering
\includegraphics[width=1\textwidth]{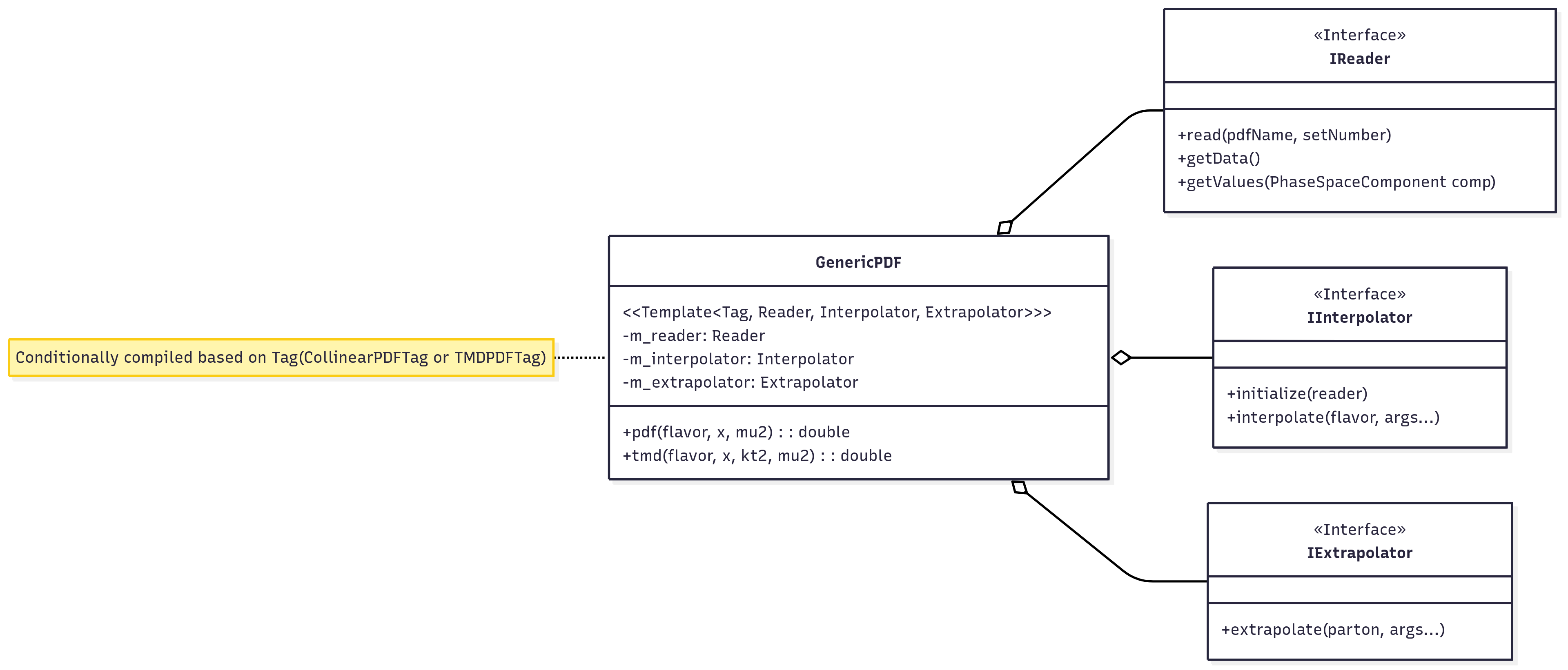}
\caption{UML class diagram of the \texttt{GenericPDF} class, showcasing its template parameters, internal members, public methods, and associations with the \texttt{IReader}, \texttt{IInterpolator}, and \texttt{IExtrapolator} interfaces.}
\label{fig:GenericPDF}
\end{figure}

The \texttt{GenericPDF} class is a fundamental component of the \texttt{PDFxTMDLib} library, designed to evaluate both cPDFs and TMDs. As illustrated in Figure~\ref{fig:GenericPDF}, it is implemented as a template class parameterized by \texttt{Tag}, \texttt{Reader}, \texttt{Interpolator}, and \texttt{Extrapolator}. The \texttt{Tag} parameter specifies the distribution type, either \texttt{CollinearPDFTag} for cPDFs or \texttt{TMDPDFTag} for TMDs, enabling conditional compilation to modify the class’s behavior and optimize performance for each case. The remaining parameters, \texttt{Reader}, \texttt{Interpolator}, and \texttt{Extrapolator}, define pluggable strategies for reading PDF grid data, interpolating between grid points, and extrapolating beyond grid boundaries, respectively.

Internally, \texttt{GenericPDF} encapsulates three private members: \texttt{m\_reader} of type \texttt{Reader}, \texttt{m\_interpolator} of type \texttt{Interpolator}, and \texttt{m\_extrapolator} of type \texttt{Extrapolator}. These members leverage the specified strategies to perform the necessary computations. The class exposes two public methods for PDF evaluation: \texttt{pdf(flavor, x, mu2)}, which computes the collinear PDF for a given parton \texttt{flavor}, momentum fraction \texttt{x}, and squared energy scale \texttt{mu2}, and \texttt{tmd(flavor, x, kt2, mu2)}, which evaluates the TMD by additionally incorporating the squared transverse momentum \texttt{kt2}. Both methods return a \texttt{double} value representing the distribution at the specified kinematic point.

The functionality of \texttt{GenericPDF} relies on three interfaces, as depicted in Figure~\ref{fig:GenericPDF}: \texttt{IReader}, \texttt{IInterpolator}, and \texttt{IExtrapolator}. The \texttt{IReader} interface provides methods for accessing PDF data, including \texttt{read(pdfName, setNumber)} to load a specific PDF set, \texttt{getData()} to retrieve the raw data, and \texttt{getValues(Phase-Space-Component comp)} to obtain values for a given phase-space component. The \texttt{IInterpolator} interface defines \texttt{initialize(reader)} to configure the interpolator with a reader instance and \texttt{interpolate(flavor, args...)} to compute interpolated values for a specified \texttt{flavor} and variable arguments. The \texttt{IExtrapolator} interface offers \texttt{extrapolate(parton, args...)} to extend evaluations beyond the grid limits for a given \texttt{parton}. The use of variadic template is one of the important extension points, that allows this approach to be extensible to higher dimensions beyond 2D or 3D of cPDFs or TMDs. 

By integrating these interfaces through its template parameters, \texttt{GenericPDF} achieves a modular and extensible architecture. This design allows users to supply custom implementations of the reader, interpolator, and extrapolator, modifying the evaluation process to specific needs while maintaining a consistent interface. The conditional compilation based on the \texttt{Tag} parameter further enhances flexibility, enabling the class to efficiently handle both cPDFs and TMDs within the same framework, with potential for future extensions to more complex distribution types.

\subsection{Grid Data Formats and File Handling}

\texttt{PDFxTMDLib} maintains compatibility with \texttt{lhagrid1} file formats of 
\texttt{LHAPDF} while introducing a new file format for TMDs which is an extension 
of the \texttt{lhagrid1} format to address the lack of standardization in the TMD 
community. We name the new file format \textit{lhagrid\_tmd1}. This extended format 
maintains structural similarity to \texttt{lhagrid1} but accommodates the additional 
dimensionality of TMDs by organizing data in grids of $x$, $k_t^2$, $\mu^2$, and 
flavors. For implementation simplicity, the current version utilizes a single 
subgrid structure, in contrast to the multiple subgrids supported in the 
\texttt{lhagrid1} format.

At the present stage, \texttt{PDFxTMDLib} provides validated support for a subset of 
TMD sets, specifically the parton-branching (PB) family:
\textit{PB-LO-HERAI+II-2020-set1}, 
\textit{PB-LO-HERAI+II-2020-set2}, 
\textit{PB-NLO+QED-HERAI+II-set1}, 
\textit{PB-NLO+QED-HERAI+II-set2}, 
\textit{PB-NLO-HERAI+II-2023-set2-qs=0.74}, and 
\textit{PB-NLO-HERAI+II-2023-set2-qs=1.04}. 
For these sets, the exact grid data files have been converted into the 
\textit{lhagrid\_tmd1} format using set-specific conversion tools, and validated by 
direct numerical comparison with \texttt{TMDLib}, as shown in \cref{validation} and 
\cref{fig:performance:TMD}, where excellent agreement is observed.

At present, \texttt{PDFxTMDLib} does \emph{not} provide a general mechanism for 
generating or converting arbitrary \texttt{TMDLib} grids into the 
\textit{lhagrid\_tmd1} format. Grid conversion and validation are currently limited to the PB sets listed above, for which the grid structure and interpolation prescriptions are well defined. Support for additional TMD formats will require dedicated, set-specific development and validation.

In addition, \texttt{PDFxTMDLib} maintains partial support for some TMD sets of 
\texttt{TMDLib} through specialized \texttt{IReader} and \texttt{IInterpolator} 
implementations for the \texttt{allflavorUpdf} format. However, due to the lack of a standard grid in \texttt{TMDLib}, this support is partial. For these 
other TMD sets, \texttt{TMDLib} remains the only choice, as it directly uses author-provided implementations that guarantee exact reproduction and correct handling of set-specific theoretical assumptions. 

The goal of \texttt{PDFxTMDLib} is therefore not to replace \texttt{TMDLib}, but to provide a unified and extensible framework that enables standardized storage and access to TMD grids where possible, while lowering the technical barrier for adding new TMD sets in the future. The library's support for TMD sets will expand progressively, with updates published published at \textit{www.pdfxtmdlib.org}.

As it is obvious, the structure of \texttt{PDFxTMDLib} makes it possible approach leads to be a general PDF framework library in which \texttt{LHAPDF} and \texttt{TMDLib} can be conceptualized as specialized implementations within the broader \texttt{PDFxTMDLib} architecture. Researchers can extend support to custom file formats by implementing appropriate \texttt{IReader}, \texttt{IInterpolator}, and \texttt{IExtrapolator} subclasses, accommodating specific requirements without modifying the overall structure of the library.

One other additional improvement offered by \texttt{PDFxTMDLib} is the extension of uncertainty and QCD coupling calculations to the TMD domain. While these features have been standard in collinear PDF libraries, \texttt{PDFxTMDLib} introduces them to TMD calculations for the first time, enabling more precise phenomenological studies.

One additional improvement offered by \texttt{PDFxTMDLib} is the extension of uncertainty and QCD coupling functionality to the TMD domain. While these features have long been standard in collinear PDF libraries, \texttt{PDFxTMDLib} introduces them to TMD analyses for the first time, enabling diagnostic and comparative studies of uncertainties directly at the distribution level. It should be noted, however, that these uncertainty estimates correspond to variations among the members of a TMD or PDF set and are not intended to represent uncertainties of physical cross sections. For cross-section calculations, all individual set members must be propagated through the full phenomenological computation. In this respect, the implementation in \texttt{PDFxTMDLib} follows the same philosophy as \texttt{LHAPDF} and \texttt{TMDplotter}, where uncertainty information is exposed at the level of the distributions themselves.

Despite the fact that many improvements and extensions are utilized in \texttt{PDFxTMDLib}, but many implementation details are just the ones of \texttt{LHAPDF} library. In fact we do not reinvent the wheel as far as possible in order the code to be reliable and correct. However, our effort was that to improve the code readability and performance, in order to be both fast and reliable. Hence, we suggest readers who are interested in the detailed physics underlying these calculations refer to the reference \cite{LHAPDF6}. However, it is of importance to discuss the standardized filesystem hierarchy for grid and metadata files. In order the grid files to be processed in \texttt{PDFxTMDLib}, they must adhere to the following file system hierarchy:
\begin{verbatim}
PDFxTMD_PATH/<setname>/<setname>_<nnnn>.dat
PDFxTMD_PATH/<setname>/<setname>.info
\end{verbatim}

In this structure, $<nnnn>$ denotes a four-digit member identifier. For instance, `CT18NLO\_0003.dat' represents the third member of the `CT18NLO' PDF set. The central metadata for each set is stored in a YAML-formatted info file bearing the set name.

To accommodate diverse installation environments, \texttt{PDFxTMDLib} implements a configurable search path mechanism. Users can specify custom search paths by editing the `config.yaml' configuration file, located at platform-specific paths:
\begin{itemize}
\item Windows: \texttt{C:\textbackslash{}ProgramData\textbackslash{}PDFxTMDLib}
\item Linux/Unix: \texttt{\textasciitilde{}/.PDFxTMDLib}
\end{itemize}

Search paths are defined using YAML syntax:
\begin{verbatim}
paths:
- /home/PDFxTMDSets1
- /home/PDFxTMDSets2
\end{verbatim}

The library implements a cascading search algorithm that examines the current binary directory, system-wide installation directories, i.e. (`/usr/local/share/PDFxTMDLib' on Linux/Unix or `C:\textbackslash ProgramData\textbackslash PDFxTMDLib' on Windows), and finally these custom paths. 

For datasets with non-standard organization, or format, \texttt{PDFxTMDLib} offers two integration approaches: (1) restructuring the dataset to conform to the library's conventions, or (2) implementing a custom reader, interpolator class that handles the specific formats.
\section{Numerical Validation and Performance of PDFxTMDLib against LHAPDF and TMDLib}
\label{validation}

In this section, we validate the numerical accuracy and performance of \texttt{PDFxTMDLib} through calculations of cross sections for the Drell-Yan process within the collinear factorization framework using the PYTHIA Monte Carlo event generator~\cite{Sjostrand:2007gs}, and also comparing results between cPDFs, and TMDs generated by \texttt{PDFxTMDLib} against \texttt{LHAPDF}, and \texttt{TMDLib}, respectively. It should be mentioned that to perform cross section calculations using PYTHIA, we have extended this Monte Carlo event generator to support \texttt{PDFxTMDLib}. Additionally, in order to ensure reproducibility of our results, we make the modified code publicly available at \url{https://pdfxtmdlib.org/downloads/}.

Finally, before presenting results, we specify the hardware and software environment used for all performance measurements:
\begin{itemize}
\item \textbf{CPU:} Intel(R) Core(TM) i7-9700 CPU @ 3.00GHz (8 cores)
\item \textbf{Operating System:} Linux Mint 21 (Vanessa)
\item \textbf{Compiler:} g++ (Ubuntu 11.4.0-1ubuntu1~22.04) 11.4.0
\end{itemize}

\subsection{Drell-Yan Process Simulations with PDFxTMDLib and LHAPDF}

The Drell-Yan process, is one of the important processes for investigation of both cPDFs, and TMDs, therefore we choose this process for to investigate the validity of \texttt{PDFxTMDLib}. In order to do this calculation, as it is mentioned before, we employ the \texttt{Pythia8} event generator~\cite{Sjostrand:2007gs} to simulate the Drell-Yan process and performing comparison against the \texttt{LHAPDF} library. The focus of this calculation is both performance, and the precision. In fact, it is crucially important for a library to be fast in addition to precise to be used in scientific domain.

We configured \texttt{Pythia8} to simulate proton-proton collisions at a center-of-mass energy of $13$ TeV, targeting the production of $Z$ bosons decaying into muon pairs (Z $\to \mu^+ \mu^-$). The simulation generated $50$ million events, with phase-space constraints enforcing an invariant mass of the dilepton system between 50 and 150 GeV and a transverse momentum exceeding 20 GeV. Two PDF configurations were tested: \texttt{LHAPDF6} with the \texttt{MSHT20nlo\_as120} set and \texttt{PDFxTMDLib} with a compatible setup. Initial-state radiation (ISR) was enabled, while final-state radiation (FSR) and multiparton interactions (MPI) were disabled to isolate the PDFs' contributions. The simulation code, written in C++ and integrated with \texttt{Pythia8}, tracked execution time and computed differential cross sections for the dilepton system's invariant mass, rapidity, and transverse momentum.

The computational performance of \texttt{PDFxTMDLib} demonstrates a modest but notable improvement over \texttt{LHAPDF}. The total execution time for 50 million events was 162,263 seconds with \texttt{PDFxTMDLib}, compared to 171,935 seconds with \texttt{LHAPDF}, yielding a reduction of approximately 5.6\% in execution time. This efficiency gain can be attributed to optimized algorithms within \texttt{PDFxTMDLib} for PDF evaluations, which reduce computational overhead during event generation. Such improvements are particularly valuable in large-scale simulations where processing time is a limiting factor. Therefor, one can see that \texttt{PDFxTMDLib} despite many layers of abstractions still has great performance and even shows better performance compared to \texttt{LHAPDF}.

\begin{figure}[htbp]
\centering
\includegraphics[width=0.49\textwidth]{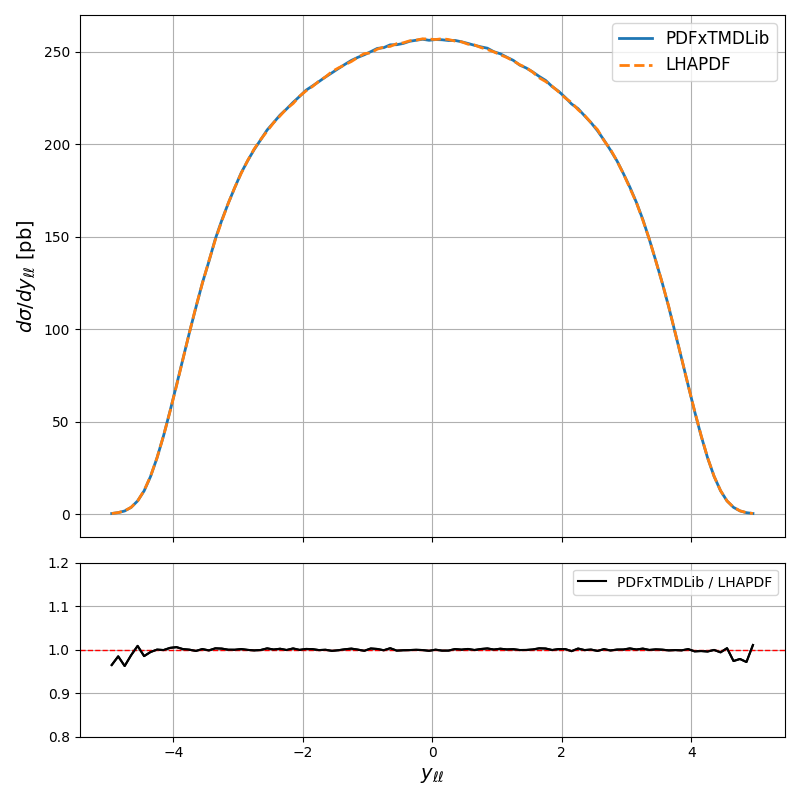}
\includegraphics[width=0.49\textwidth]{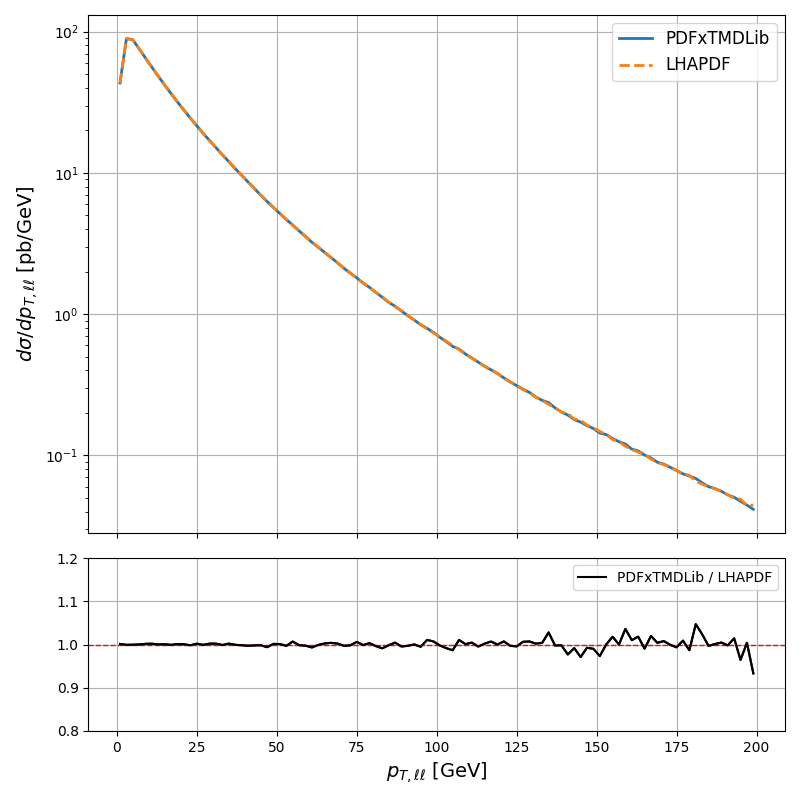}
\includegraphics[width=0.5\textwidth]{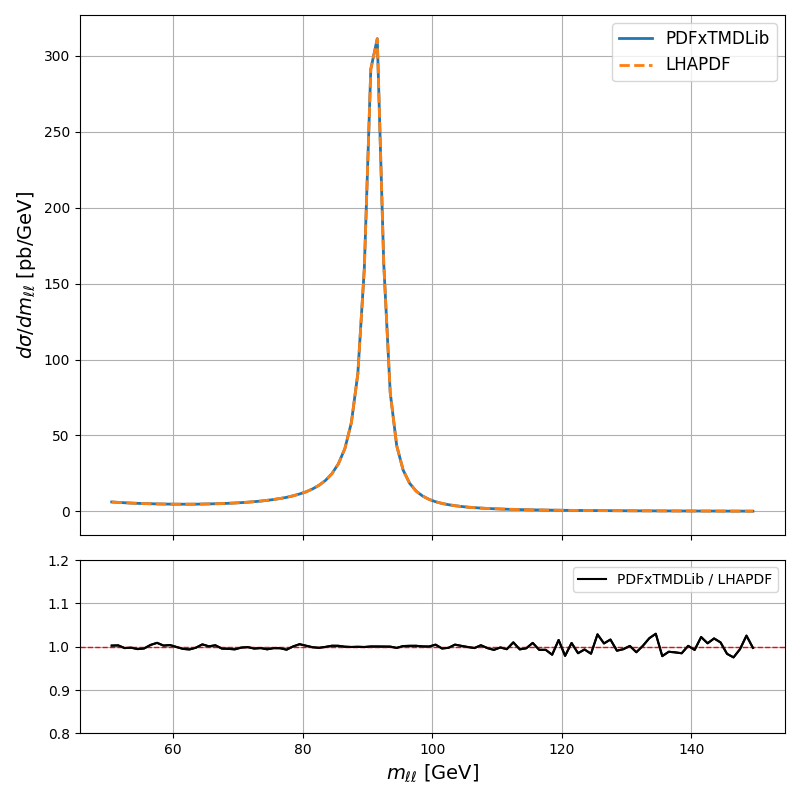}
\caption{	Comparison of differential cross sections for the Drell--Yan process
	simulated with \texttt{PYTHIA} using \texttt{PDFxTMDLib} (solid lines)
	and \texttt{LHAPDF} (dashed lines) with the MSHT20nlo\_as120 PDF set.
	The upper panels show the absolute distributions, while the lower panels
	show the ratio $\mathrm{PDFxTMDLib}/\mathrm{LHAPDF}$; the horizontal line
	indicates unity.
	From top left to bottom, the observables shown are the dilepton rapidity
	$y_{\ell\ell}$, transverse momentum $p_{T,\ell\ell}$, and invariant mass
	$m_{\ell\ell}$.}
\label{fig:DrellYan}
\end{figure}

To assess the validity of \texttt{PDFxTMDLib}, we compare the total and differential cross sections with those obtained using \texttt{LHAPDF}. The total Drell–Yan cross section is measured to be $1.695 \times 10^{-6}$ mb in both libraries, with statistical uncertainties of $1.453 \times 10^{-10}$ mb, demonstrating perfect agreement. Differential cross sections, presented as histograms of the dilepton invariant mass, rapidity, and transverse momentum, also show excellent consistency between the two implementations. These comparisons are displayed in Figure~\ref{fig:DrellYan}, where the overlaid distributions confirm the alignment across all kinematic variables. This agreement indicates that \texttt{PDFxTMDLib} provides a reliable alternative to \texttt{LHAPDF}.

It is worth noting, however, that a direct performance comparison with \texttt{TMDLib} is not straightforward, as \texttt{TMDLib} consists of multiple independently developed implementations corresponding to different TMD sets, each exhibiting distinct computational characteristics. Consequently, a uniform benchmarking comparison across its various components is non-trivial. For this reason, we do not attempt such a comparison here, although interested readers may pursue it independently. In the following subsection, we examine the validity of \texttt{PDFxTMDLib} in more detail by comparing both its collinear PDFs (cPDFs) and TMDs against the corresponding results from \texttt{LHAPDF} and \texttt{TMDLib}.

\subsection{Validation of PDFxTMDLib for cPDFs and TMDs}

\begin{figure}[htbp]
\centering
\includegraphics[width=0.48\textwidth]{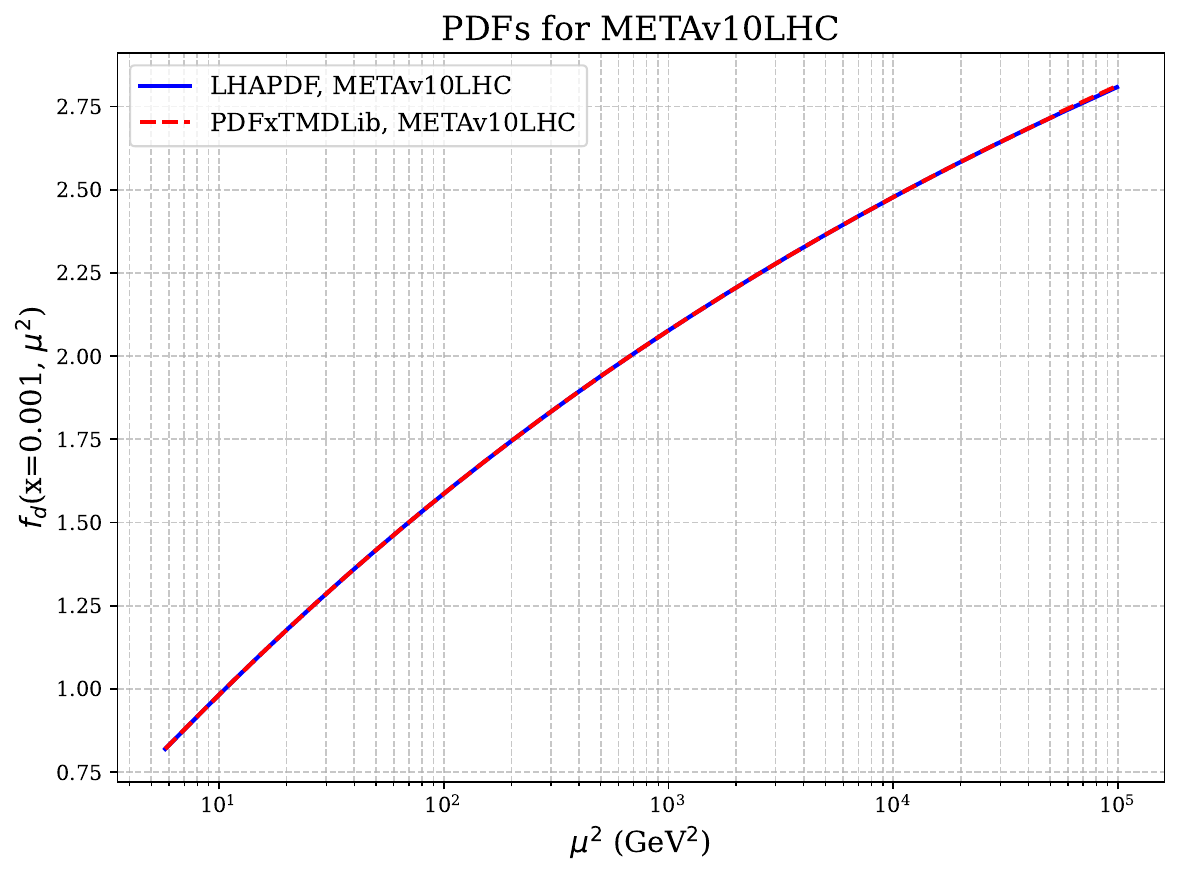}
\includegraphics[width=0.48\textwidth]{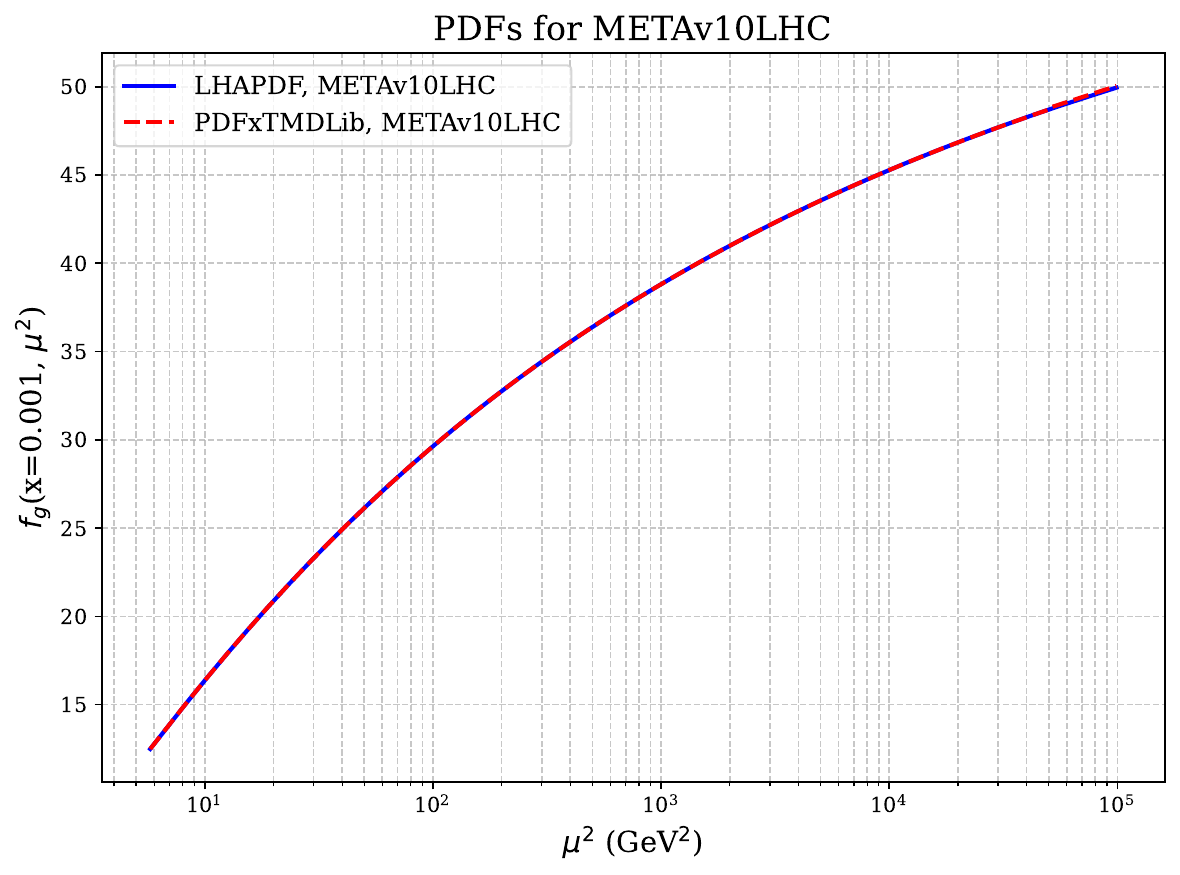}
\includegraphics[width=0.48\textwidth]{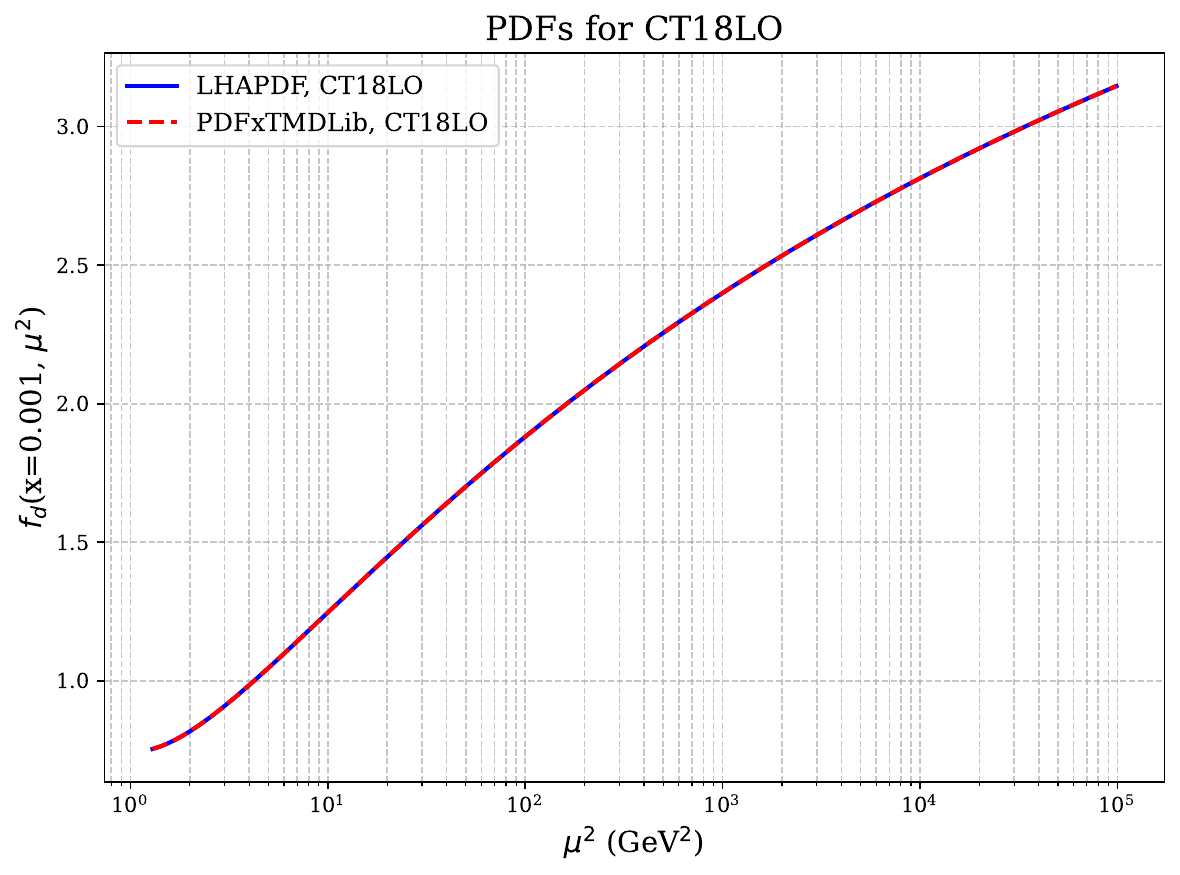}
\includegraphics[width=0.48\textwidth]{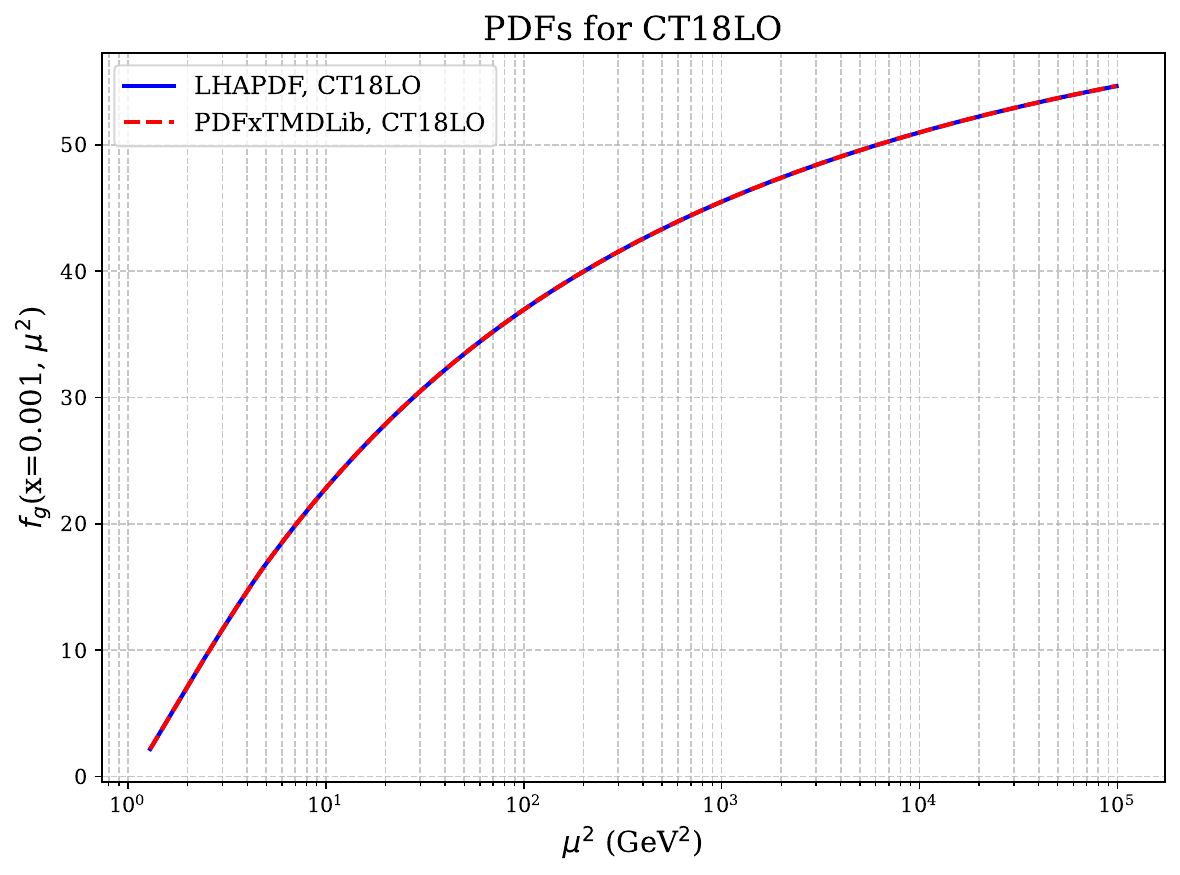}
\includegraphics[width=0.48\textwidth]{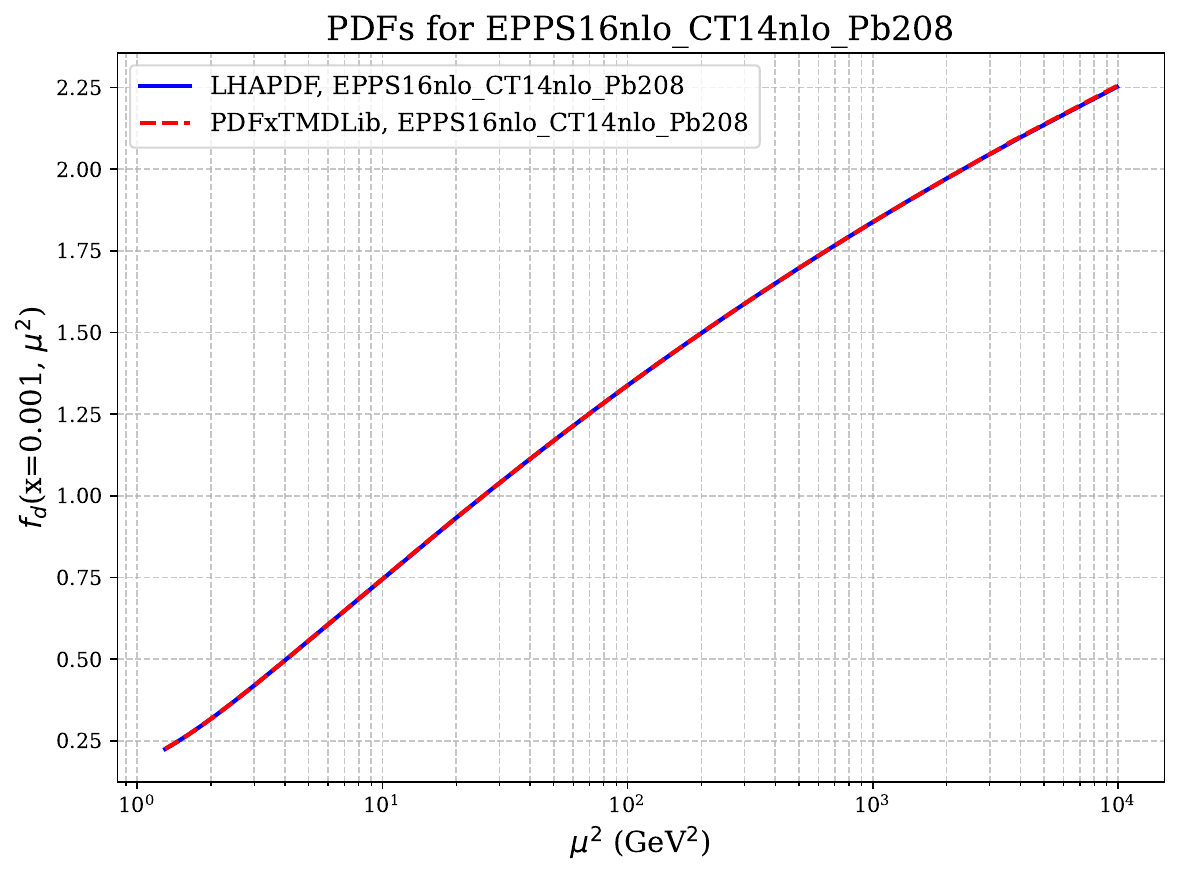}
\includegraphics[width=0.48\textwidth]{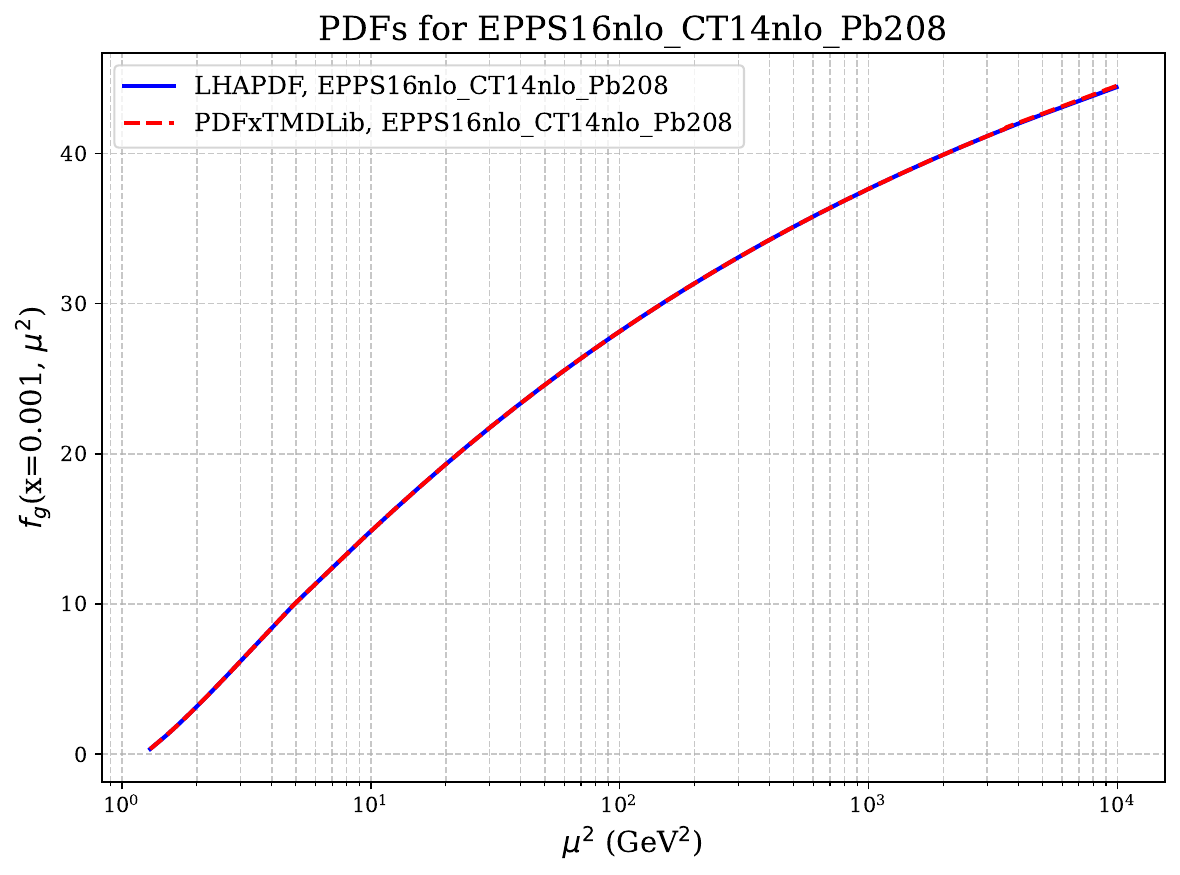}
\caption{Comparison of collinear parton distribution functions(cPDFs) for the down quark (left) and gluon (right) at $x=0.001$ as a function of $\mu^2$, using \texttt{PDFxTMDLib} (solid lines) and \texttt{LHAPDF} (dashed lines) for the \texttt{METAv10LHC} (top), \texttt{CT18LO} (middle), and \texttt{EPPS16nlo\_CT14nlo\_Pb208} (bottom) PDF sets. The agreement between the two libraries validates the implementation in \texttt{PDFxTMDLib}.}
\label{fig:performance-CPDF}
\end{figure}

\begin{figure}[htbp]
\centering
\includegraphics[width=0.48\textwidth]{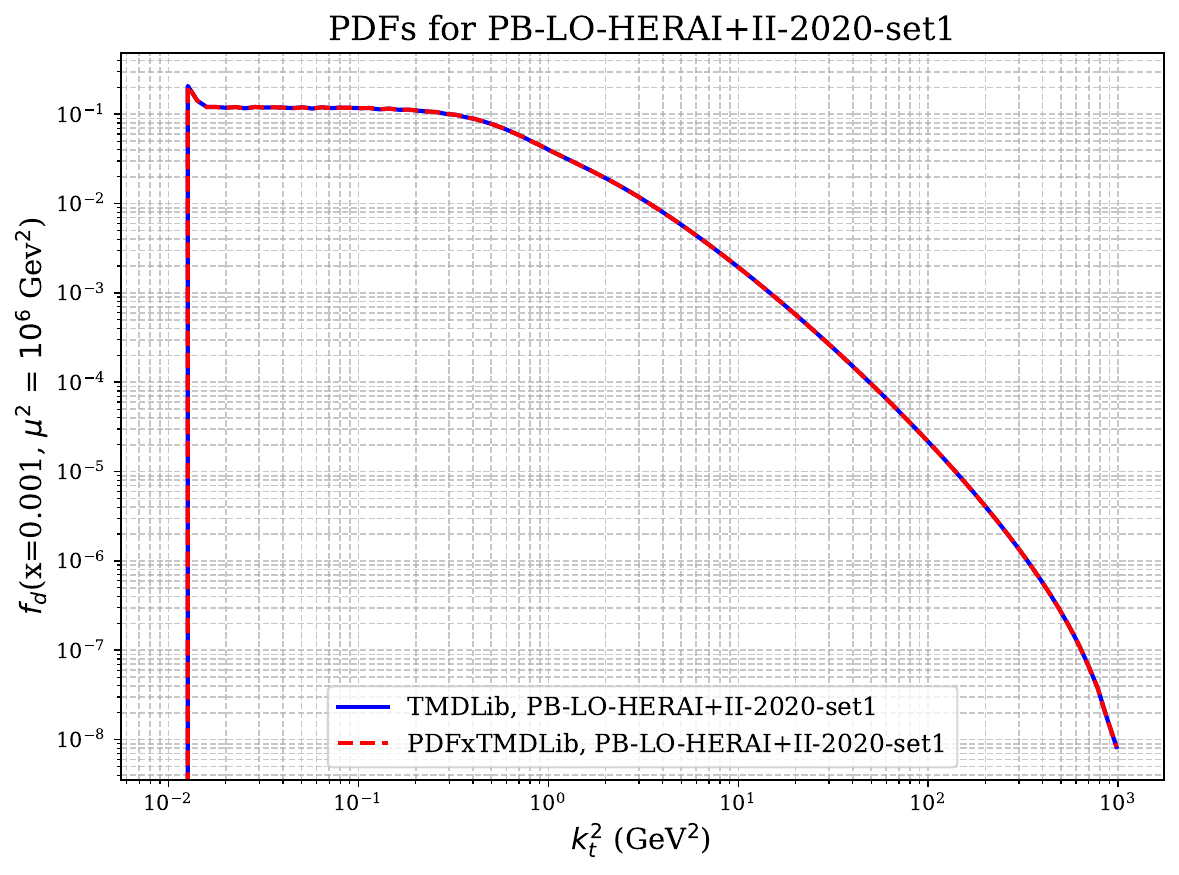}
\includegraphics[width=0.48\textwidth]{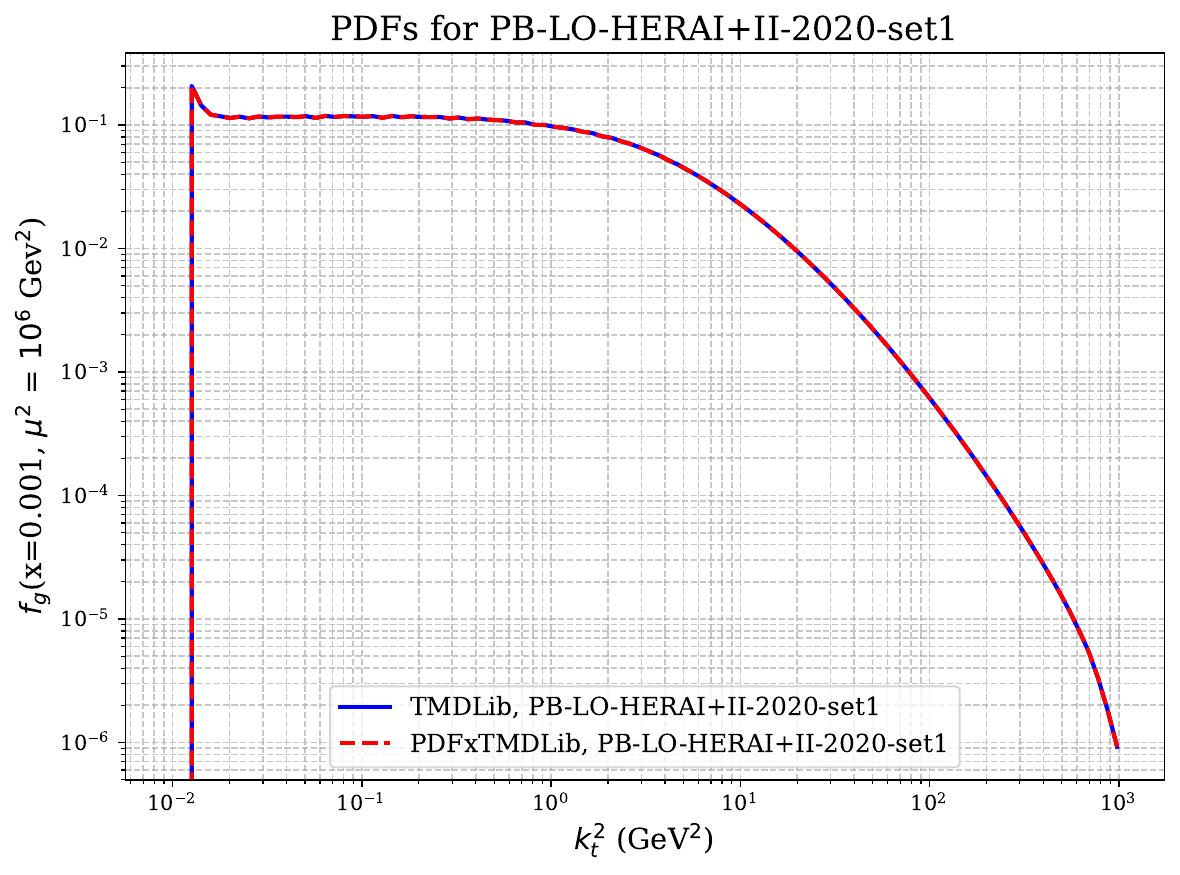}
\includegraphics[width=0.48\textwidth]{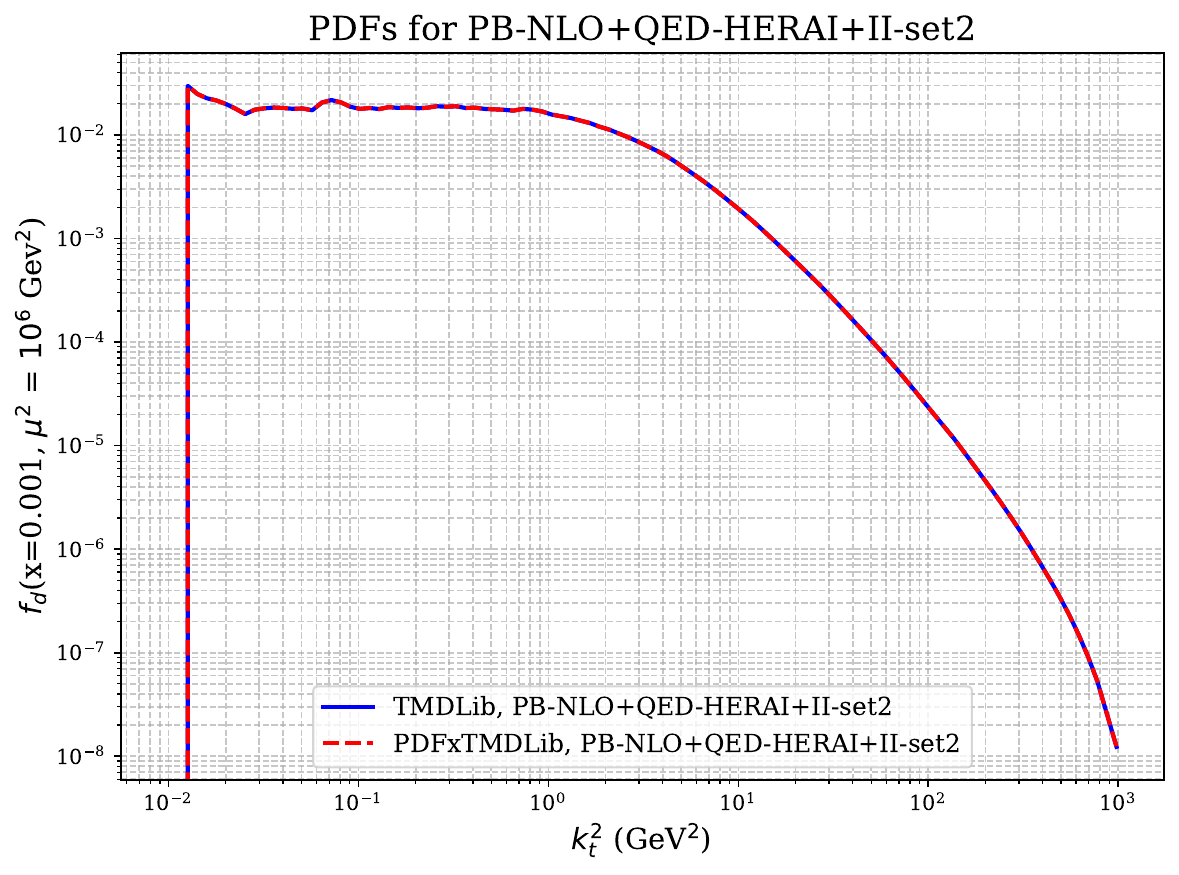}
\includegraphics[width=0.48\textwidth]{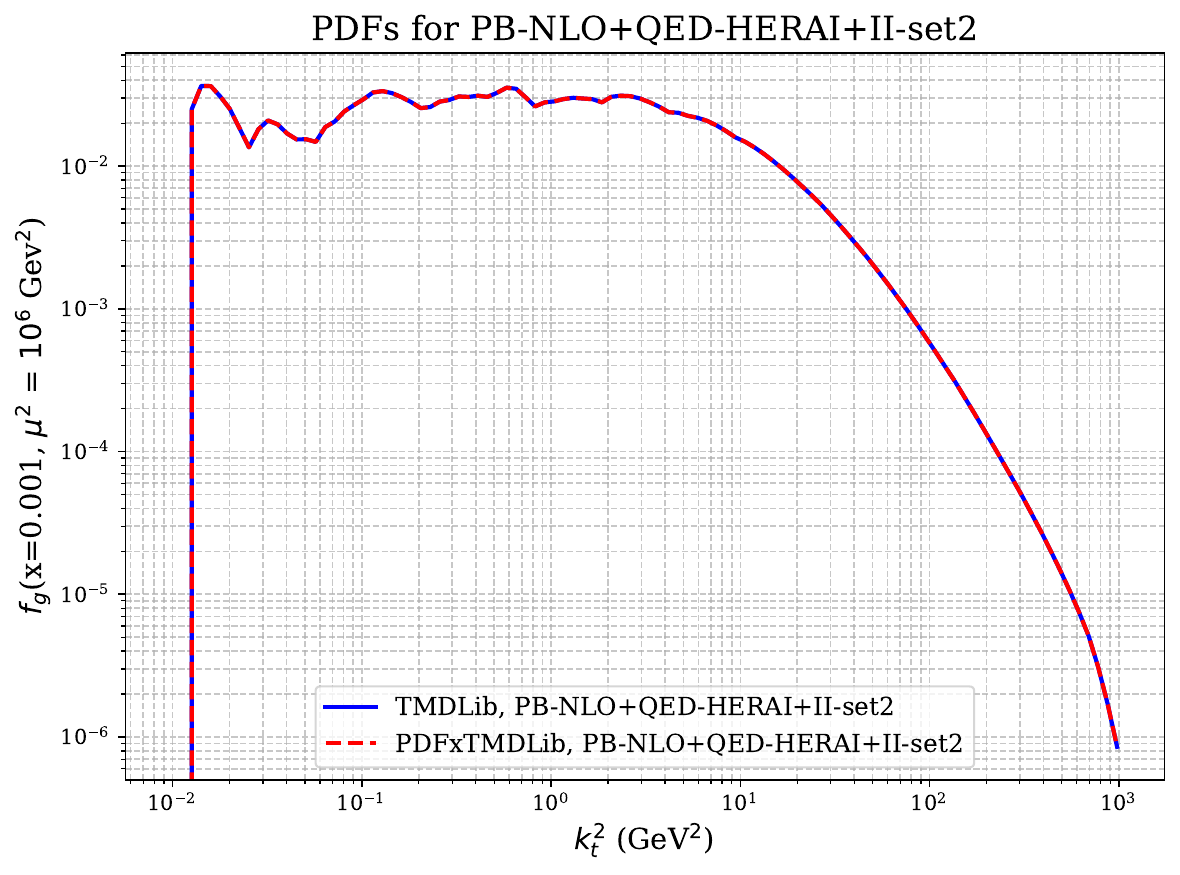}
\includegraphics[width=0.48\textwidth]{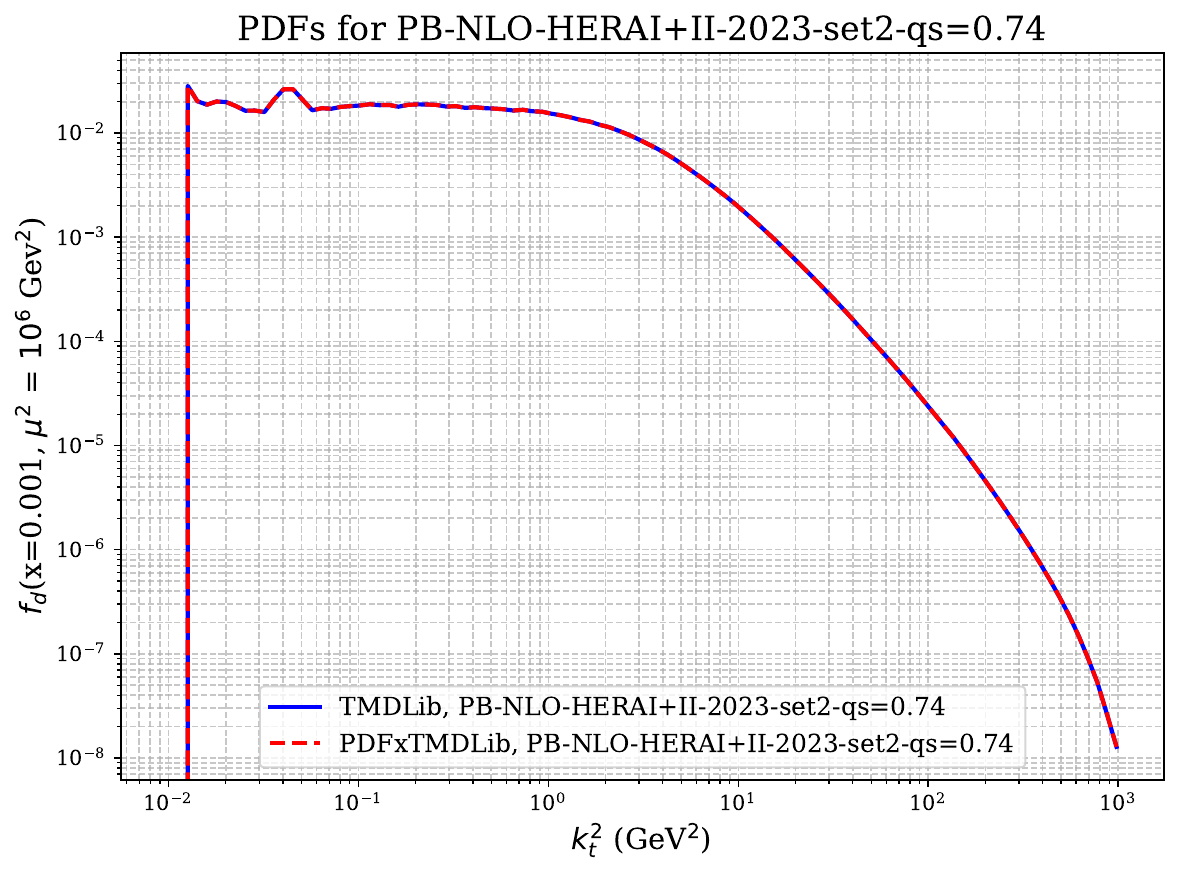}
\includegraphics[width=0.48\textwidth]{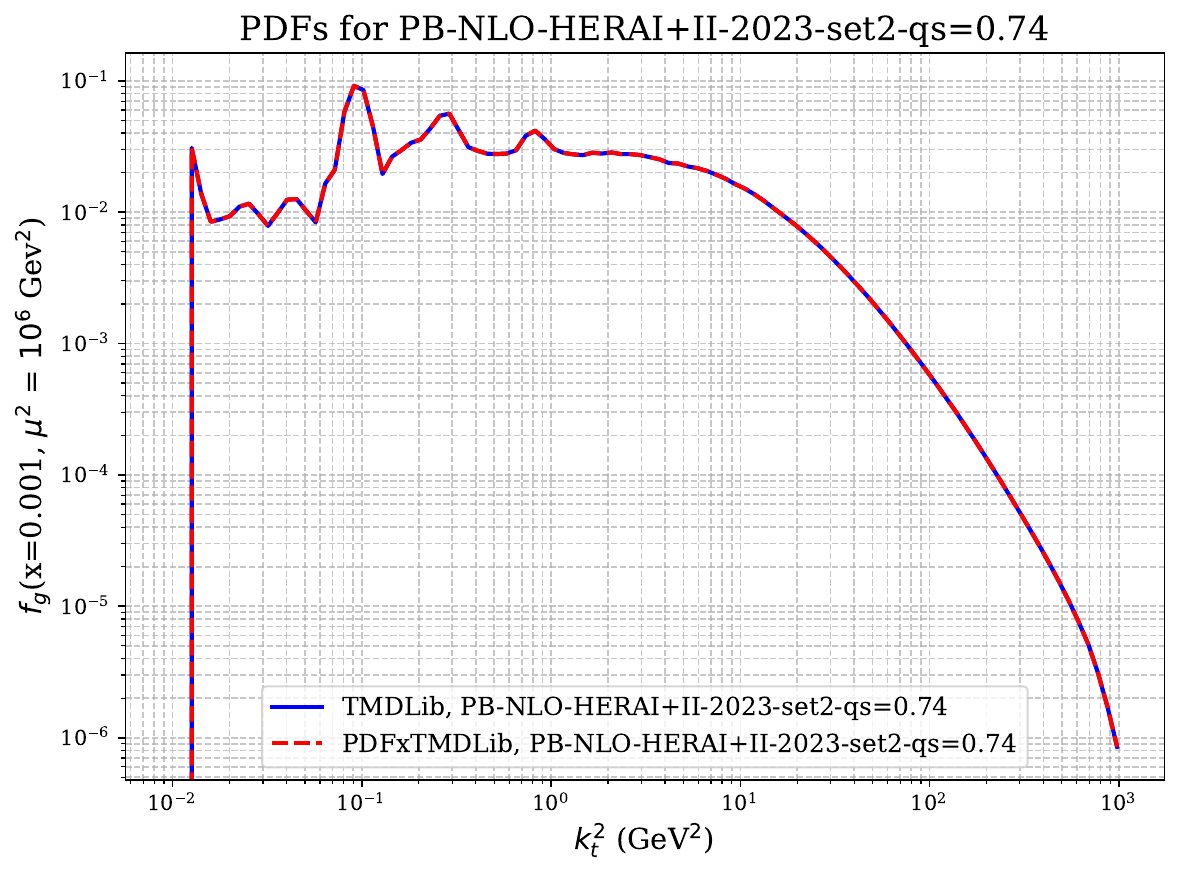}
\caption{Comparison of transverse momentum dependent PDFs (TMDs) for the down quark (left) and gluon (right) at $x=0.001$ and $k_T=1000$ (units assumed, e.g., GeV) as a function of $\mu^2$, using \texttt{PDFxTMDLib} (solid lines) and TMDLib (dashed lines) for the \texttt{PB-LO-HERAI+II-2020-set1} (top), \texttt{PB-NLO+QED-HERAI+II-set2} (middle), and \texttt{PB-NLO-HERAI+II-2023-set2-qs=0.74} (bottom) TMD sets. The matching curves confirm the correctness of \texttt{PDFxTMDLib}'s TMD implementation.}
\label{fig:performance:TMD}
\end{figure}

In this subsection, we present comparisons of the cPDFs, and TMDs generated by \texttt{PDFxTMDLib} against those from \texttt{LHAPDF}, and \texttt{TMDLib}, for the gluon and down quark distributions. To perform this analysis, we calculate cPDFs at $x=0.001$ over the range of $\mu^2$ from the minimum to the maximum allowed by each PDF set. As shown in Figure~\ref{fig:performance-CPDF}, we perform this comparison for the following PDF sets: \texttt{METAv10LHC}, \texttt{CT18LO}, and \texttt{EPPS16nlo\_CT14nlo\_Pb208}. The results generated by \texttt{PDFxTMDLib} completely match those of the \texttt{LHAPDF} library. Similarly, for TMDs, we calculate distributions at $x=0.001$ and a fixed transverse momentum $k_t=1000$ (units assumed, e.g., GeV), over the range of $\mu^2$, using the sets \texttt{PB-LO-HERAI+II-2020-set1}, \texttt{PB-NLO-HERAI+II-2023-set2-qs=0.74}, and \texttt{PB-NLO+QED-HERAI+II-set2}, as illustrated in Figure~\ref{fig:performance:TMD}. The results from \texttt{PDFxTMDLib} fully agree with those from \texttt{TMDLib}, confirming the correctness and validity of this library.

	\section{Installation and Usage Guide}
\label{installationAndUsage}
This section provides instructions for building \texttt{PDFxTMDLib} and integrating it into research workflows.
The source code is publicly available on GitHub at \url{https://github.com/Raminkord92/PDFxTMD}, and the latest versions can also be downloaded from the project webpage at \url{https://pdfxtmdlib.org/downloads/} (e.g.\ \url{https://pdfxtmdlib.org/downloads/PDFxTMD-main.zip}).
For any questions regarding the installation or usage of the library, users can contact us by email or via the official GitHub discussions page: \url{https://github.com/Raminkord92/PDFxTMD/discussions}.

\subsection{Building and Installing PDFxTMDLib}

\subsubsection{Prerequisites}
Before installing \texttt{PDFxTMDLib}, ensure your system meets these requirements:
\begin{itemize}
	\item C++17 compatible compiler (GCC 8+, Clang 7+, MSVC 2019+)
	\item CMake 3.14 or newer
	\item For Windows: Microsoft Visual Studio 2019 or newer
\end{itemize}

\subsubsection{Build Process}
The library uses a standard CMake build process. Execute the following commands in a terminal (Linux/macOS) or Command Prompt/PowerShell (Windows):

\begin{lstlisting}[style=bash]
	cmake -S . -B build -DCMAKE_INSTALL_PREFIX=/path/to/install -DCMAKE_BUILD_TYPE=Release
	cmake --build build
\end{lstlisting}

\subsubsection{Installation}
To install the library, execute:
\begin{lstlisting}[style=bash]
	cmake --build build --target install
\end{lstlisting}

This installs headers, libraries, and CMake configuration files under the chosen prefix. If no prefix is specified, CMake will use its default install location (which may be a system directory such as \texttt{/usr/local} on Linux/macOS).

\subsection{Integration Methods}

\subsubsection{CMake Integration (Recommended)}
For projects using CMake, \texttt{PDFxTMDLib} can be easily integrated by adding these lines to your CMakeLists.txt file:

\begin{lstlisting}[style=CMake]
	find_package(PDFxTMDLib REQUIRED)
	target_link_libraries(your-target-name PDFxTMD::PDFxTMDLib)
\end{lstlisting}

\subsubsection{Direct Compilation}
For projects not using CMake, you can link directly with the compiler:

\begin{lstlisting}[style=bash]
	# Linux/macOS with GCC/Clang
	g++ -std=c++17 your_source.cpp -lPDFxTMDLib -o your_executable
	
	# Windows with MSVC (from Developer Command Prompt)
	cl your_source.cpp /std:c++17 PDFxTMDLib.lib
\end{lstlisting}

\subsection{API Overview}
\texttt{PDFxTMDLib} offers a robust API designed for both high-level convenience and low-level flexibility, suitable to a wide range of scientific applications. This section presents key usage patterns through practical examples, progressing from basic interfaces to advanced customization. Complete reference implementations are available in the project's GitHub repository: \url{https://github.com/Raminkord92/PDFxTMD/blob/main/examples/}.

\subsection{High-Level Interface for PDF Sets}
The \texttt{PDFSet} template class provides a unified high-level interface for PDF calculations, uncertainty analysis, and metadata access. It abstracts internal complexities, such as loading PDF set members, reading metadata file, and select appropriate reader, interpolator, and extrapolator, as a result of this making it ideal for most users.

\subsubsection{Collinear PDF Calculations}
The high-level interface for cPDFs uses the \texttt{PDFSet} class specialized with \texttt{CollinearPDFTag}. Below is an example of instantiating a collinear PDF set and evaluating the gluon PDF:

\begin{lstlisting}[style=cpp, breaklines=true, basicstyle=\ttfamily\footnotesize]
	#include <PDFxTMDLib/PDFSet.h>
	#include <iostream>
	int main()
	{
		// Instantiate a PDFSet for collinear distributions
		PDFxTMD::PDFSet<PDFxTMD::CollinearPDFTag> cpdfSet("MSHT20nlo_as120");
		// Access the central member ( index 0)
		auto central_pdf = cpdfSet[0];
		// Define kinematics
		double x = 0.1;     // Longitudinal momentum fraction
		double mu2 = 10000; // Factorization scale squared
		// Evaluate PDF for gluon
		double gluon_pdf = central_pdf->pdf(PDFxTMD::PartonFlavor ::g, x, mu2);
		std ::cout << "Gluon PDF at x=" << x << ", mu2 =" << mu2 << " GeV2 : " << gluon_pdf
		<< std ::endl;
		return 0;
	}
\end{lstlisting}

\subsubsection{TMD PDF Calculations}
For TMDs, the \texttt{PDFSet} class is specialized with \texttt{TMDPDFTag}. The following example evaluates the up-quark TMD, incorporating the transverse momentum parameter $k_t^2$:

\begin{lstlisting}[style=cpp, breaklines=true, basicstyle=\ttfamily\footnotesize]
#include <PDFxTMDLib/PDFSet.h>
#include <iostream>
int main()
{
	// Instantiate a PDFSet for TMD distributions
	PDFxTMD::PDFSet<PDFxTMD::TMDPDFTag> tmdSet("PB-LO-HERAI+II-2020-set2");
	// Access the central member ( index 0)
	auto central_tmd = tmdSet[0];
	// Define kinematics
	double x = 0.01;  // Longitudinal momentum fraction
	double kt2 = 10;  // Transverse momentum squared
	double mu2 = 100; // Factorization scale squared
	// Evaluate TMD for up quark
	double up_tmd = central_tmd->tmd(PDFxTMD::PartonFlavor::u, x, kt2, mu2);
	std::cout << "Up - quark TMD at x=" << x << ", kT2=" << kt2 << ", mu2=" << mu2
	<< " GeV2 : " << up_tmd << std::endl;
	return 0;
}
\end{lstlisting}

\subsubsection{Uncertainty and Correlation Analysis}
\texttt{PDFxTMDLib} automates uncertainty and correlation calculations, selecting the appropriate statistical method (e.g., Hessian or Monte Carlo replicas) based on PDF set metadata. This applies to both collinear and TMD PDFs:

\begin{lstlisting}[style=cpp, breaklines=true, basicstyle=\ttfamily\footnotesize]
	// Calculate PDF uncertainty at default confidence level
	PDFxTMD::PDFUncertainty uncertainty = cpdfSet.Uncertainty(
	PDFxTMD::PartonFlavor::g, x, mu2);
	
	std::cout << "xg = " << uncertainty.central
	<< " + " << uncertainty.errplus
	<< " - " << uncertainty.errminus << std::endl;
	
	// Calculate uncertainty at 90% confidence level
	PDFxTMD::PDFUncertainty uncertainty_90 = cpdfSet.Uncertainty(
	PDFxTMD::PartonFlavor::g, x, mu2, 90.0);
	
	// Calculate correlation between gluon and up quark
	double correlation = cpdfSet.Correlation(
	PDFxTMD::PartonFlavor::g, x, mu2,
	PDFxTMD::PartonFlavor::u, x, mu2);
	
	std::cout << "Correlation between g and u: " << correlation << std::endl;
\end{lstlisting}
\textbf{Important note:}
The uncertainty values returned by \texttt{Uncertainty(...)} refer only to the PDF or TMD distributions and not to physical cross sections.

\subsubsection{Factory Interfaces for Individual PDF Members}
For applications requiring only specific PDF members without uncertainty calculations, factory interfaces offer an efficient alternative:

\begin{lstlisting}[style=cpp, breaklines=true, basicstyle=\ttfamily\footnotesize]
	#include <PDFxTMD/GenericCPDFFactory.h>
	#include <iostream>
	
	int main() {
		// Create a factory
		auto cpdf_factory = PDFxTMD::GenericCPDFFactory();
		
		// Create a single PDF member
		auto cpdf = cpdf_factory.mkCPDF("MMHT2014lo68cl", 0);
		
		// Evaluate PDF directly
		double x = 0.001, mu2 = 100;
		double gluon_pdf = cpdf.pdf(PDFxTMD::PartonFlavor::g, x, mu2);
		std::cout << "Gluon PDF: " << gluon_pdf << std::endl;
		
		return 0;
	}
\end{lstlisting}

\subsubsection{QCD Coupling Calculations}
The library supports calculating the strong coupling constant $\alpha_s(\mu^2)$ via two methods: using \texttt{PDFSet} or \texttt{CouplingFactory}:

\begin{lstlisting}[style=cpp, breaklines=true, basicstyle=\ttfamily\footnotesize]
	PDFSet<CollinearPDFTag> cpdfSet("MSHT20nlo_as120");
	double alpha_sCPDF = cpdfSet.alphasQ2(mu2);
	
	auto couplingFactory = CouplingFactory();
	auto coupling = couplingFactory.mkCoupling("MMHT2014lo68cl");
	double alpha_s = coupling.AlphaQCDMu2(mu2);
\end{lstlisting}

\subsection{Advanced Usage Patterns}
For advanced users, \texttt{PDFxTMDLib} supports customization through template specialization and provides type aliases, i.e. \texttt{CollinearPDF}, and \texttt{TMDPDF} for convenience. These two aliases are convenient for most of the works. \texttt{CollinearPDF} use bicubic interpolator, with continuation extrapolator, which are equivalent to default \texttt{LHAPDF} selection choice. While \texttt{TMDPDF} use trilinear interplator (\texttt{TMDLib} default choice for \texttt{PB}-family sets), and zero extraploator:

\begin{lstlisting}[style=cpp, breaklines=true, basicstyle=\ttfamily\footnotesize]
	// Custom PDF implementation
	using ExtrapolatorType = CErrExtrapolator;
	using ReaderType = CDefaultLHAPDFFileReader;
	using InterpolatorType = CLHAPDFBilinearInterpolator<ReaderType>;
	using PDFType = GenericPDF<CollinearPDFTag, ReaderType, InterpolatorType, ExtrapolatorType>;
	PDFType cpdf("MMHT2014lo68cl", 0);
	
	// Using type aliases
	CollinearPDF cpdfPDF("MMHT2014lo68cl", 0);
	TMDPDF tmdPDF("PB-LO-HERAI+II-2020-set2", 0);
\end{lstlisting}

\subsection{Python Interface and Integration}

Finally, \texttt{PDFxTMDLib} provides Python bindings that allow to utilize most of the features of this library easily in python language:

\subsubsection{Installation and Basic Usage}

The Python interface can be installed using pip:

\begin{lstlisting}[style=bash]
	pip install pdfxtmd
\end{lstlisting}

The interface mirrors the C++ API structure:

\begin{lstlisting}[style=cpp,  breaklines=true, basicstyle=\ttfamily\footnotesize]
	import pdfxtmd
	import numpy as np
	import matplotlib.pyplot as plt
	
	# Initialize a collinear PDF set
	cpdf_set = pdfxtmd.CPDFSet("CT18NLO")
	
	# Access metadata
	std_info = cpdf_set.getStdPDFInfo()
	print(f"PDF set: {std_info.SetDesc}({std_info.NumMembers} members)")
	
	# Evaluate PDFs for visualization
	x_values = np.logspace(-4, -1, 100)  # x range from 10^-4 to 10^-1
	mu2 = 1000  # GeV^2
	kt2 = 1  # GeV^2
	
	# Calculate gluon PDFs and uncertainties
	gluon_pdfs = [cpdf_set[0].pdf(pdfxtmd.PartonFlavor.g, x, mu2) for x in x_values]
	uncertainties = [cpdf_set.Uncertainty(pdfxtmd.PartonFlavor.g, x, mu2) for x in x_values]
	
	# Extract upper and lower uncertainty bands for PDFs
	upper_band = [g + u.errplus for g, u in zip(gluon_pdfs, uncertainties)]
	lower_band = [g - u.errminus for g, u in zip(gluon_pdfs, uncertainties)]
	
	# Initialize a TMD set
	tmd_set = pdfxtmd.TMDSet("PB-LO-HERAI+II-2020-set2")
	# Access metadata
	std_info = tmd_set.getStdPDFInfo()
	print(f"TMD set: {std_info.SetDesc}({std_info.NumMembers} members)")
	
	# Calculate gluon TMDs and uncertainties
	gluon_tmds = [tmd_set[0].tmd(pdfxtmd.PartonFlavor.g, x, kt2, mu2) for x in x_values]
	tmd_uncertainties = [tmd_set.Uncertainty(pdfxtmd.PartonFlavor.g, x, kt2, mu2) for x in x_values]
	
	# Extract upper and lower uncertainty bands for TMDs
	tmd_upper_band = [g + u.errplus for g, u in zip(gluon_tmds, tmd_uncertainties)]
	tmd_lower_band = [g - u.errminus for g, u in zip(gluon_tmds, tmd_uncertainties)]
\end{lstlisting}

Complete examples, including uncertainty analysis, PDF calculation, and visualization code, are available in the project repository: \url{https://github.com/Raminkord92/PDFxTMD/blob/main/examples} directory.
\section{Conclusion}
\label{conclusion}
In this paper, we have presented \texttt{PDFxTMDLib}, a C++ library developed to address the computational needs of both cPDFs and TMDs in high energy physics. By integrating modern C++ design principles, such as the Curiously Recurring Template Pattern, and type erasure, also modular interfaces, \texttt{PDFxTMDLib} offers a high-performance and extensible framework that overcomes several limitations of existing tools like \texttt{LHAPDF} and \texttt{TMDLib}. Its unified approach facilitates efficient handling of cPDFs and TMDs, while its flexible architecture supports custom implementations and future extensions, even to higher-dimensional distributions.

Numerical validation, including Drell-Yan process simulations with PYTHIA and direct comparisons with \texttt{LHAPDF} and \texttt{TMDLib}, demonstrates that \texttt{PDFxTMDLib} delivers consistent and accurate results, alongside modest performance improvements in specific scenarios. The library further enhances phenomenological research by introducing novel features for TMDs, such as uncertainty quantification at distribution level and QCD coupling calculations, which were previously unavailable in a standardized form. Additionally, we also introduce a new \textit{lhagrid\_tmd1} file format for TMDs which is an extension of \textit{lhagrid} file format to facilitate and standardize calculations for TMDs.

\texttt{PDFxTMDLib} stands as a practical tool for researchers, supporting PDF-related computations and enabling adaptable workflows through its C++ and Python interfaces. Its design ensures compatibility with evolving theoretical advancements, contributing to the study of hadron structure and high energy collision dynamics. Looking ahead, potential developments include broader support for additional PDF and TMD sets, algorithmic optimizations, and extensions to distributions like double parton distribution functions (DPDFs).

\section*{Data Availability Statement}
No Data associated in the manuscript.

\bibliographystyle{ieeetr}

\end{document}